\documentclass[11pt,a4paper,twocolumn]{article}
\pdfoutput=1
\usepackage[utf8]{inputenc}
\usepackage{amsmath}
\usepackage{amssymb}
\usepackage{graphicx}
\usepackage[hidelinks,hypertexnames=false]{hyperref}
\usepackage{geometry}
\usepackage{tikz}
\usetikzlibrary{shapes,arrows,positioning,patterns,calc}
\usepackage{float}
\usepackage{pgfplots}
\usepackage{pgfplotstable}
\pgfplotsset{compat=1.18}
\usepackage{booktabs}
\usepackage{multirow}
\usepackage{xcolor}
\usepackage{placeins}
\usepackage{balance}
\usepackage{enumitem}
\setlist{nosep, leftmargin=*, topsep=2pt, partopsep=0pt}
\usepackage[activate={true,nocompatibility},final,tracking=true,factor=1100,stretch=10,shrink=10]{microtype}
\usepackage{adjustbox}
\usepackage{capt-of}
\usepackage{subcaption}
\makeatletter
\renewcommand{\section}{\@startsection{section}{1}{\z@}%
  {-2.5ex \@plus -0.5ex \@minus -.2ex}{1.2ex \@plus.2ex}%
  {\normalfont\Large\bfseries\raggedright}}
\renewcommand{\subsection}{\@startsection{subsection}{2}{\z@}%
  {-2.0ex\@plus -0.5ex \@minus -.2ex}{0.8ex \@plus .2ex}%
  {\normalfont\large\bfseries\raggedright}}
\renewcommand{\subsubsection}{\@startsection{subsubsection}{3}{\z@}%
  {-1.5ex\@plus -0.5ex \@minus -.2ex}{0.6ex \@plus .2ex}%
  {\normalfont\normalsize\bfseries\raggedright}}
\makeatother
\newcounter{algorithm}
\renewcommand{\thealgorithm}{\arabic{algorithm}}

\renewcommand{\arraystretch}{1.05}
\title{From LLM to Silicon: RL-Driven ASIC Architecture Exploration\\
for On-Device AI Inference}

\author{
Ravindra Ganti\\
\texttt{ravindra@xgensilicon.com}\\
XgenSilicon Inc.\\
\and
Steve Xu\\
\texttt{steve@xgensilicon.com}\\
XgenSilicon Inc.
}

\date{April 2026}
\newcommand{\ProcessNodeRange}{3nm to 28nm}

\newcommand{\NumProcessNodes}{7}
\newcommand{\BestNode}{3nm}
\newcommand{\WorstNode}{28nm}
\newcommand{\BestMesh}{41$\times$42}

\newcommand{\BestCores}{1,722}

\newcommand{\BestTokS}{29,809}
\newcommand{\BestPPA}{0.974}
\newcommand{\BestPowerMW}{51,366}
\newcommand{\BestPowerW}{51}
\newcommand{\BestPerfGOps}{466,364}
\newcommand{\BestAreaMMtwo}{648}
\newcommand{\WorstPowerMW}{3,780}

\newcommand{\WorstPerfGOps}{9,744}
\newcommand{\WorstAreaMMtwo}{3,545}
\newcommand{\WorstPPA}{1.019}

\newcommand{\PerfRatio}{47.85}

\newcommand{\AreaReductionFactor}{5.47}
\newcommand{\PPAImprovementFactor}{1.05}
\newcommand{\TotalWeightsMetric}{14.96 GB}
\newcommand{\TotalParameters}{8.03B}
\newcommand{\ModelPrecision}{FP16}
\newcommand{\ModelName}{Llama 3.1 8B}
\newcommand{\GraphOperators}{7,489}
\newcommand{\TotalInstructions}{597M}
\newcommand{\MaxSequenceLength}{2,048}
\newcommand{\OptimizationMode}{high-performance}

\newcommand{\MaxEpisodesThreeNm}{4,613}

\begin{document}

\maketitle

\begin{abstract}
We present an RL-driven compiler that jointly optimizes ASIC architecture, memory hierarchy, and workload partitioning for AI inference across \ProcessNodeRange{}. The design space is formulated as a single Markov Decision Process with mixed discrete-continuous actions and a unified Power-Performance-Area (PPA) objective. Soft Actor-Critic (SAC) with Mixture-of-Experts gating explores the joint space of mesh topology, per-core microarchitecture, and operator placement. We validate on two workloads: \ModelName{} \ModelPrecision{} (high-performance mode, \BestTokS{} tok/s at \BestNode{}) and SmolVLM (low-power mode, $<$13\,mW at all nodes, 10\,MHz). Across \NumProcessNodes{} process nodes, the RL automatically adapts mesh sizes and per-tile configurations---including heterogeneous FETCH, VLEN, and memory allocation---without node-specific manual retuning.
\end{abstract}

\section{Introduction}

The deployment of large-scale neural networks on custom AI accelerators requires careful co-optimization of hardware architecture and software compilation strategies. Traditional ASIC design flows require months of manual RTL coding and verification. Recent work has explored automated optimization using reinforcement learning~\cite{mirhoseini2017device,gao2018tpu}, but existing methods optimize individual components in isolation rather than jointly optimizing the system stack from model input to silicon output.

Our compiler uses a unified RL-based optimization over a 2D mesh of Tensor Compute Cores (TCCs)---RISC-V cores with custom vector/tensor extensions---jointly determining mesh topology, per-TCC microarchitecture (FETCH, VLEN, memory sizes), workload partitioning, and NoC configuration. The key insight is that \textit{joint} optimization over these coupled dimensions yields better PPA than optimizing them independently.

Our contributions are:
\begin{enumerate}
    \item \textbf{Joint MDP formulation:} A 73-dimensional state (52-dim SAC subset) and 30-dimensional continuous action space with SAC+MoE policy that co-optimizes mesh topology, per-TCC parameters, and operator placement.
    \item \textbf{Heterogeneous per-TCC derivation:} Workload-adaptive FETCH, VLEN, and memory allocation per tile, reducing power without sacrificing throughput.
    \item \textbf{Multi-workload, multi-node validation:} Evaluation on Llama 3.1 8B (high-performance) and SmolVLM (low-power) across \NumProcessNodes{} process nodes (\ProcessNodeRange{}) with automated artifact-to-paper data pipeline.
\end{enumerate}

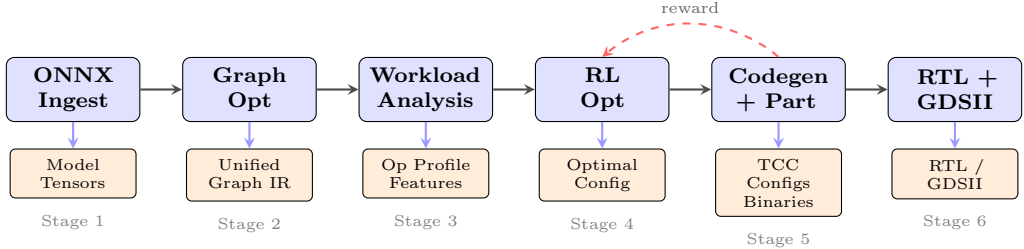
\begin{figure*}[!htbp]
\centering
\begin{tikzpicture}[
    node distance=0.15cm,
    stage/.style={rectangle, draw, fill=blue!12, text width=1.5cm, text centered, rounded corners=3pt, minimum height=0.85cm, font=\scriptsize\bfseries},
    artifact/.style={rectangle, draw, fill=orange!15, text width=1.4cm, text centered, rounded corners=2pt, minimum height=0.6cm, font=\tiny},
    arrow/.style={->, >=stealth, thick, color=black!70},
    darrow/.style={->, >=stealth, thick, dashed, color=red!60},
    label/.style={font=\tiny, color=black!60}
]
    \node [stage] (s1) {ONNX\\Ingest};
    \node [stage, right=0.55cm of s1] (s2) {Graph\\Opt};
    \node [stage, right=0.55cm of s2] (s3) {Workload\\Analysis};
    \node [stage, right=0.55cm of s3] (s4) {RL\\Opt};
    \node [stage, right=0.55cm of s4] (s5) {Codegen\\+ Part};
    \node [stage, right=0.55cm of s5] (s6) {RTL +\\GDSII};

    \draw [arrow] (s1) -- (s2);
    \draw [arrow] (s2) -- (s3);
    \draw [arrow] (s3) -- (s4);
    \draw [arrow] (s4) -- (s5);
    \draw [arrow] (s5) -- (s6);

    \node [artifact, below=0.35cm of s1] (a1) {Model\\Tensors};
    \node [artifact, below=0.35cm of s2] (a2) {Unified\\Graph IR};
    \node [artifact, below=0.35cm of s3] (a3) {Op Profile\\Features};
    \node [artifact, below=0.35cm of s4] (a4) {Optimal\\Config};
    \node [artifact, below=0.35cm of s5] (a5) {TCC Configs\\Binaries};
    \node [artifact, below=0.35cm of s6] (a6) {RTL /\\GDSII};

    \draw [arrow, color=blue!40] (s1) -- (a1);
    \draw [arrow, color=blue!40] (s2) -- (a2);
    \draw [arrow, color=blue!40] (s3) -- (a3);
    \draw [arrow, color=blue!40] (s4) -- (a4);
    \draw [arrow, color=blue!40] (s5) -- (a5);
    \draw [arrow, color=blue!40] (s6) -- (a6);

    \draw [darrow] (s5.north) to[out=140,in=40] node[above,label] {reward} (s4.north);

    \node [below=0.08cm of a1, font=\tiny, color=gray] {Stage 1};
    \node [below=0.08cm of a2, font=\tiny, color=gray] {Stage 2};
    \node [below=0.08cm of a3, font=\tiny, color=gray] {Stage 3};
    \node [below=0.08cm of a4, font=\tiny, color=gray] {Stage 4};
    \node [below=0.08cm of a5, font=\tiny, color=gray] {Stage 5};
    \node [below=0.08cm of a6, font=\tiny, color=gray] {Stage 6};
\end{tikzpicture}
\caption{End-to-end compilation pipeline. The RL optimization loop (Stage 4) receives workload features from Stage 3 and drives codegen/partitioning in Stage 5, which returns PPA reward signals (dashed arrow). Each stage emits intermediate artifacts used by downstream stages and for reproducibility.}
\label{fig:pipeline}
\end{figure*}

Figure~\ref{fig:pipeline} illustrates the design flow from model ingestion through tape-out-ready ASIC output.

\section{Related Work}

\subsection{Neural Network Compilation}
TVM~\cite{chen2018tvm} introduced AutoTVM for automated operator-kernel tuning through template-guided search; however, schedule templates must be authored manually for each hardware target, and the framework does not jointly optimize architecture-level parameters. Ansor~\cite{lee2021ansor} extends this to template-free search but remains limited to single-operator tuning without cross-operator partitioning awareness. MLIR~\cite{larus2020mlir} provides a multi-level IR that simplifies progressive lowering but exposes no built-in PPA-aware optimization loop. TensorFlow XLA~\cite{xla} and PyTorch Glow~\cite{rotem2018glow} fuse and schedule subgraphs for existing accelerators but cannot retarget across process nodes or co-optimize hardware parameters. TensorRT~\cite{chen2018tensorrt} focuses on NVIDIA GPU inference and does not generalize to custom ASIC design spaces.

\subsection{Reinforcement Learning for Hardware Design}
Mirhoseini et al.~\cite{mirhoseini2017device} applied RL to device placement in distributed systems, demonstrating that policy-gradient methods can outperform expert placements. Their follow-up work~\cite{mirhoseini2020chip} extended RL to chip floorplanning with graph neural network state encoders. Gao et al.~\cite{gao2018tpu} used RL for TPU datapath optimization. However, these approaches optimize single design phases (placement or datapath) in isolation rather than jointly optimizing architecture, memory hierarchy, and workload partitioning. Our method unifies these into a single MDP with mixed discrete-continuous actions.

\subsection{Auto-Tuning and Search}
Bayesian optimization~\cite{rasmussen2006gaussian} has been applied to hyperparameter tuning with Gaussian-process surrogate models. Genetic algorithms~\cite{whitley1994genetic} and simulated annealing~\cite{kirkpatrick1983optimization} provide derivative-free global search but lack the ability to learn from sequential state transitions. These methods scale poorly when the design space combines mesh topology, per-core memory, and partitioning decisions. In contrast, RL-based search exploits temporal structure in the MDP and reuses learned value estimates across episodes.

\subsection{Neural Architecture Search}
NAS~\cite{zoph2016neural,real2019regularized} has demonstrated automated architecture discovery, but targets model topology (layer types, connections) rather than hardware-software co-design. Hardware-aware NAS variants~\cite{wu2019fbnet} incorporate latency predictors but still treat the hardware as fixed. Our work takes the complementary view: the model is given, and the hardware+compiler stack is optimized.

\subsection{Previous Work on Hardware-Aware Compilation}
Our previous work~\cite{ganti2025hardware} introduced hardware-aware neural network compilation with learned optimization for RISC-V accelerators, focusing on instruction-level optimization and register allocation for individual cores. The current work extends this foundation along four axes: (1) multi-core mesh architecture optimization with heterogeneous per-core parameter allocation, (2) operation-level partitioning across compute cores, (3) process-node retargeting across 3nm to 28nm, and (4) end-to-end automation from AI inference models to GDSII with no manual retuning.

Table~\ref{tab:related_comparison} positions our approach relative to prior systems on key capability dimensions.

\begin{table*}[!htbp]
\centering
\scriptsize
\setlength{\tabcolsep}{2pt}
\renewcommand{\arraystretch}{1.2}
\begin{tabular}{l|ccc|ccccc|cccc}
\hline
 & \multicolumn{3}{c|}{\textbf{Compiler Stack}} & \multicolumn{5}{c|}{\textbf{Optimization Features}} & \multicolumn{4}{c}{\textbf{Differentiating Capabilities}} \\
\textbf{System} & \rotatebox{70}{\textbf{Frontend/IR}} & \rotatebox{70}{\textbf{Graph Opt.}} & \rotatebox{70}{\textbf{Backend/Codegen}} & \rotatebox{70}{\textbf{RL/AutoTune}} & \rotatebox{70}{\textbf{Quantization}} & \rotatebox{70}{\textbf{Memory Plan}} & \rotatebox{70}{\textbf{PPA Model}} & \rotatebox{70}{\textbf{Target}} & \rotatebox{70}{\textbf{Joint HW+SW}} & \rotatebox{70}{\textbf{Multi-Core}} & \rotatebox{70}{\textbf{Op-Level Part.}} & \rotatebox{70}{\textbf{Node Retarget}} \\
\hline
TVM~\cite{chen2018tvm}               & \checkmark & \checkmark & \checkmark & \checkmark & \checkmark & \texttimes & \texttimes & CPU/GPU   & \texttimes & \texttimes & \texttimes & \texttimes \\
Ansor~\cite{lee2021ansor}             & \checkmark & \checkmark & \checkmark & \checkmark & \texttimes & \texttimes & \texttimes & CPU/GPU   & \texttimes & \texttimes & \texttimes & \texttimes \\
XLA~\cite{xla}                        & \checkmark & \checkmark & \checkmark & \texttimes & \checkmark & \checkmark & \texttimes & TPU/GPU   & \texttimes & \checkmark & \texttimes & \texttimes \\
MLIR~\cite{larus2020mlir}             & \checkmark & \checkmark & \checkmark & \texttimes & \texttimes & \texttimes & \texttimes & Multi     & \texttimes & \texttimes & \texttimes & \texttimes \\
Glow~\cite{rotem2018glow}             & \checkmark & \checkmark & \checkmark & \texttimes & \checkmark & \checkmark & \texttimes & CPU/Accel & \texttimes & \texttimes & \texttimes & \texttimes \\
TensorRT~\cite{chen2018tensorrt}      & \checkmark & \checkmark & \checkmark & \texttimes & \checkmark & \checkmark & \texttimes & GPU       & \texttimes & \texttimes & \texttimes & \texttimes \\
Mirhoseini~\cite{mirhoseini2020chip}  & \texttimes & \texttimes & \texttimes & \checkmark & \texttimes & \texttimes & \texttimes & ASIC      & \texttimes & \checkmark & \texttimes & \texttimes \\
Timeloop~\cite{parashar2019timeloop}  & \texttimes & \texttimes & \texttimes & \texttimes & \texttimes & \checkmark & \checkmark & ASIC      & \texttimes & \texttimes & \texttimes & \texttimes \\
\textbf{Ours}                         & \checkmark & \checkmark & \checkmark & \checkmark & \checkmark & \checkmark & \checkmark & ASIC      & \checkmark & \checkmark & \checkmark & \checkmark \\
\hline
\end{tabular}
\vspace{1pt}
\caption{Capability comparison with prior ML compiler systems. \textit{Compiler Stack}: standard compiler infrastructure. \textit{Optimization Features}: RL/AutoTune (automated search), Quantization (FP16/INT8), Memory Plan (buffer/cache optimization), PPA Model (power-performance-area), Target (hardware platform). \textit{Differentiating Capabilities}: features unique to our approach.}
\label{tab:related_comparison}
\end{table*}

\section{Methodology}

\subsection{Problem Formulation}

We formulate hardware-software co-optimization as a Markov Decision Process (MDP) where:
\begin{itemize}
    \item \textbf{State $s_t$:} Current configuration, workload characteristics, and per-core metrics
    \item \textbf{Action $a_t$:} Parameter adjustments (mesh dimensions, per-core parameters, partitioning ratios)
    \item \textbf{Reward $r_t$:} PPA score with constraint penalties
    \item \textbf{Policy $\pi_\theta(a|s)$:} Neural network mapping states to action distributions
\end{itemize}

\subsection{State Representation}

The full state vector $\mathbf{s} \in \mathbb{R}^{73}$ captures the complete system state; the SAC actor operates on a 52-dimensional optimized subset. Table~\ref{tab:state_features} provides the breakdown.

\begin{table*}[!htbp]
\centering
\footnotesize
\setlength{\tabcolsep}{3pt}
\renewcommand{\arraystretch}{1.05}
\begin{tabular}{p{3.0cm}ccp{9.2cm}}
\hline
\textbf{Category} & \textbf{Idx Range} & \textbf{Dims} & \textbf{Representative Features} \\
\hline
\textbf{Workload} & 0--4 & 5 & Instruction count, ILP, memory intensity, vector util, matmul ratio \\
\textbf{Configuration} & 5--25 & 21 & Mesh size, fetch/STANUM/VLEN, DMEM/WMEM/IMEM, NoC width, ports, node \\
\textbf{Partitioning} & 26--28 & 3 & DMEM input/output/scratch allocation ratios \\
\textbf{Load Distribution} & 29--32 & 4 & Load variance, max/min load ratio, balance score \\
\textbf{Op Partition} & 33--36 & 4 & MatMul/Conv/general partitioning ratios \\
\textbf{Hazards} & 37--40 & 4 & Global RAW/WAR/WAW statistics \\
\textbf{Per-TCC Hazards} & 41--44 & 4 & Per-core hazard aggregates \\
\textbf{Frequency} & 45 & 1 & Clock frequency (normalized) \\
\textbf{Streaming} & 46--49 & 4 & Streaming and pipeline features \\
\textbf{PPA Observation} & 50--54 & 5 & Surrogate PPA feedback (power, perf, area, tok/s, efficiency) \\
\textbf{Workload Partition} & 55--58 & 4 & Per-TCC workload distribution statistics \\
\textbf{Precision Dist.} & 59--64 & 6 & FP32/FP16/BF16/FP8/INT8/mixed ratios \\
\textbf{Instruction Type} & 65--66 & 2 & Scalar/vector instruction ratios \\
\textbf{SC Topology} & 67--69 & 3 & Effective TCC count, avg hops, SC latency \\
\textbf{LLM Config} & 70--72 & 3 & Batch size, KV strategy, KV compression \\
\hline
\textbf{Total} & \textbf{0--72} & \textbf{73} & \textbf{Full state (SAC uses 52-dim optimized subset)} \\
\hline
\end{tabular}
\vspace{1pt}
\caption{State feature breakdown for RL optimization (73 total, 52 used by SAC actor)}
\label{tab:state_features}
\end{table*}

\subsection{Action Space}

The action space combines discrete and continuous actions. The SAC policy outputs 30 continuous action dimensions (mapped to 51-dim policy targets via quantization); 4 discrete mesh/SC deltas are sampled separately. Table~\ref{tab:action_space} provides the breakdown.

\begin{table*}[!htbp]
\centering
\footnotesize
\setlength{\tabcolsep}{3pt}
\renewcommand{\arraystretch}{1.05}
\begin{tabular}{p{3.4cm}ccp{9.3cm}}
\hline
\textbf{Action Group} & \textbf{Idx Range} & \textbf{Dims} & \textbf{Update Scope} \\
\hline
\textbf{Discrete Mesh/SC Deltas} & 0--3 & 4 & Mesh width/height and SC x/y in \{-2..+2\} (5-way one-hot each) \\
\textbf{Continuous TCC Params} & 4--18 & 15 & Fetch, STANUM, VLEN, DMEM/WMEM/IMEM, DFLIT, ports, clock, precision \\
\textbf{Memory/Load Partition} & 19--22 & 4 & DMEM input/output fractions and load-balance controls \\
\textbf{Op-Partition Controls} & 23--25 & 3 & MatMul/Conv/general operation split ratios across TCCs \\
\textbf{Streaming} & 26--27 & 2 & Input/output streaming ratio controls \\
\textbf{Workload Partition} & 28--29 & 2 & Sub-matmul partition and all-reduce fraction \\
\hline
\textbf{Total} & \textbf{0--29} & \textbf{30} & \textbf{SAC continuous dims (+ 20-dim discrete = 80-dim policy output)} \\
\hline
\end{tabular}
\vspace{1pt}
\caption{Action-space breakdown (30 continuous + 4 discrete mesh deltas)}
\label{tab:action_space}
\end{table*}

The discrete actions enable coarse-grained exploration of mesh dimensions, while continuous actions provide fine-grained parameter tuning. This hybrid approach balances exploration efficiency with optimization precision.

\textbf{Per-core vs.\ global configuration scope.} The RL agent optimizes average TCC parameters (Continuous TCC Params group in Table~\ref{tab:action_space}). A post-RL derivation step then computes \emph{per-TCC heterogeneous} values for FETCH\_SIZE, VLEN, DMEM, IMEM, and WMEM based on each tile's workload characteristics (compute load, hazard density, weight footprint). Only STANUM and the NoC-level DFLIT\_WIDTH remain uniform. The effective RL dimensionality per episode is $4 + 13 + 4 + 3 + 2 + 2 + 2 = 30$ (mesh/SC + TCC + partition + op-partition + register/NoC + streaming + workload), not $M \times N \times 13$ as per-core independent tuning would require.

This heterogeneous derivation produces per-tile configurations that can vary significantly: FETCH\_SIZE ranges 1--16 (93.8\% variation), VLEN ranges 128--2048 bits (93.8\% variation), WMEM varies by $>$30\% across tiles (see Section~\ref{sec:wmem_analysis}). Tiles hosting memory-heavy operators (attention projections, MLP layers) receive larger WMEM and wider SIMD, while tiles with lighter workloads receive smaller allocations to save area and power.

\subsection{Policy Network Architecture}

\begin{figure}[ht]
\centering
\begin{tikzpicture}[
    node distance=0.7cm, scale=0.70, every node/.style={transform shape},
    block/.style={rectangle, draw, fill=blue!15, minimum width=4.2cm, text centered, rounded corners=3pt, minimum height=0.8cm, font=\small, inner sep=4pt},
    head/.style={rectangle, draw, fill=green!15, minimum width=2.4cm, text centered, rounded corners=3pt, minimum height=0.8cm, font=\small, inner sep=4pt},
    feat/.style={rectangle, draw, fill=gray!15, minimum width=1.8cm, text centered, rounded corners=2pt, minimum height=0.6cm, font=\scriptsize, inner sep=3pt},
    arrow/.style={->, >=stealth, semithick},
]
    \node [feat] (f1) {Workload (5)};
    \node [feat, right=0.1cm of f1] (f2) {Config (21)};
    \node [feat, right=0.1cm of f2] (f3) {PPA (14)};
    \node [feat, right=0.1cm of f3] (f4) {Prec+LLM (12)};

    \node [block, below=0.5cm of f2, xshift=1cm] (concat) {Concat $\to$ $\mathbf{s} \in \mathbb{R}^{52}$ (SAC subset)};

    \draw [arrow] (f1.south) -- (concat);
    \draw [arrow] (f2.south) -- (concat);
    \draw [arrow] (f3.south) -- (concat);
    \draw [arrow] (f4.south) -- (concat);

    \node [block, below=0.4cm of concat, fill=blue!25] (hidden1) {Linear$(52\!\to\!256)$ + GELU};
    \draw [arrow] (concat) -- (hidden1);

    \node [block, below=0.35cm of hidden1, fill=blue!15] (hidden2) {Linear$(256\!\to\!256)$ + GELU};
    \draw [arrow] (hidden1) -- (hidden2);

    \node [head, below left=0.6cm and 0.6cm of hidden2, align=center] (d_head) {Softmax\\ $\mathbf{p}$ $(4\!\times\!5)$};
    \node [head, below=0.6cm of hidden2, align=center] (c_mean) {Tanh\\ $\boldsymbol{\mu}$ (30)};
    \node [head, below right=0.6cm and 0.6cm of hidden2, fill=yellow!20, align=center] (c_std) {Clamp\\ $\log\boldsymbol{\sigma}$ (30)};

    \draw [arrow] (hidden2) -- (d_head);
    \draw [arrow] (hidden2) -- (c_mean);
    \draw [arrow] (hidden2) -- (c_std);

    \node [below=0.2cm of d_head, font=\scriptsize\bfseries] {Mesh/SC $\Delta$};
    \node [below=0.2cm of c_mean, font=\scriptsize\bfseries] {TCC $\boldsymbol{\mu}$};
    \node [below=0.2cm of c_std, font=\scriptsize\bfseries] {TCC $\boldsymbol{\sigma}$};
\end{tikzpicture}
\caption{SAC actor network: $\mathbf{s} \in \mathbb{R}^{52} \to$ 2-layer MLP (256 hidden) $\to$ 80-dim output (20 discrete logits + 30 means + 30 log-stds). Actions sampled via tanh-squashed Gaussian with reparameterization.}
\label{fig:policy_network}
\end{figure}
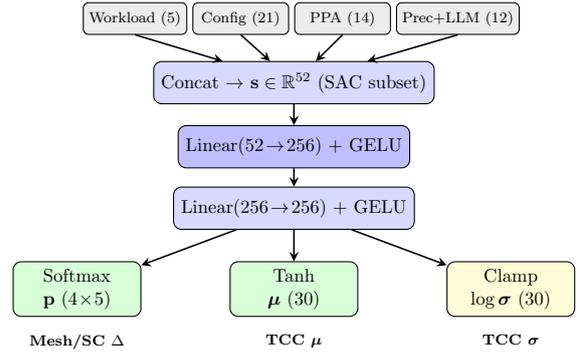

The policy network $\pi_\theta(a|s)$ uses two hidden layers followed by action-specific heads. The full state vector has 73 features; SAC operates on a 52-dimensional optimized subset. The architecture is illustrated in Figure~\ref{fig:policy_network} and mathematically defined as:

\begin{align}
\mathbf{h}_1 &= \text{GELU}(\mathbf{W}_1 \mathbf{s} + \mathbf{b}_1) \label{eq:hidden} \\
\mathbf{h}_2 &= \text{GELU}(\mathbf{W}_5 \mathbf{h}_1 + \mathbf{b}_5) \label{eq:hidden2} \\
\mathbf{p}_{\text{disc}} &= \text{softmax}(\mathbf{W}_2 \mathbf{h}_2 + \mathbf{b}_2) \label{eq:discrete_prob} \\
\mu_{\text{cont}} &= \tanh(\mathbf{W}_3 \mathbf{h}_2 + \mathbf{b}_3) \label{eq:continuous_mean} \\
\log\sigma_{\text{cont}} &= \text{clamp}(\mathbf{W}_4 \mathbf{h}_2 + \mathbf{b}_4,\, {-}20, 2) \label{eq:continuous_std}
\end{align}

where:
\begin{itemize}
    \item $\mathbf{W}_1 \in \mathbb{R}^{256 \times 52}$, $\mathbf{W}_5 \in \mathbb{R}^{256 \times 256}$: Hidden layers
    \item $\mathbf{W}_2 \in \mathbb{R}^{20 \times 256}$: Discrete head (4 mesh/SC deltas $\times$ 5 options)
    \item $\mathbf{W}_3, \mathbf{W}_4 \in \mathbb{R}^{30 \times 256}$: Continuous mean / log-std heads
\end{itemize}

The actor uses GELU activation with tanh-squashed Gaussian sampling: $a = \tanh(\mu + \sigma \odot \epsilon)$, $\epsilon \sim \mathcal{N}(0, I)$. Log-std is clamped to $[-20, 2]$ for numerical stability.

\subsubsection{Multi-Discrete Action Sampling}

For discrete actions (mesh dimensions), we sample from categorical distributions:
\begin{align}
a_{\text{mesh width}} &\sim \text{Categorical}(p_{\text{mesh width}}) \label{eq:discrete_sampling_width} \\
a_{\text{mesh height}} &\sim \text{Categorical}(p_{\text{mesh height}}) \label{eq:discrete_sampling_height}
\end{align}

For continuous actions, we sample from truncated normal distributions:
\begin{align}
a_i &\sim \mathcal{N}(\mu_i, \sigma_i^2), \text{ clipped to } [a_{\min}, a_{\max}] \label{eq:continuous_sampling}
\end{align}

\subsubsection{Adaptive Exploration}

We use epsilon-greedy exploration with adaptive decay. The base rate $d$ is auto-derived from the episode budget to reach $\epsilon_{\min}$ from $\epsilon_0$. When no feasible configurations have been discovered, the decay is slowed:
\begin{align}
\epsilon_{t+1} = \begin{cases}
\epsilon_t \times d & \text{if feasible configs found} \\
\epsilon_t \times d' & \text{otherwise, } d' > d
\end{cases} \label{eq:adaptive_exploration}
\end{align}
where $d' = 1 - (1-d) \times 0.1$ blends toward slower decay, keeping exploration high until the policy discovers feasible regions of the design space.

The exploration rate adapts based on whether feasible configurations are being discovered, allowing more exploration when stuck.

\subsection{Operation-Level Partitioning}

A key innovation is the ability to partition individual operations across multiple compute cores. For partitionable operations (matrix multiplication, convolution), we use the following procedure:
\begin{enumerate}
    \item \textbf{Determine Operation Type:} $\text{type} = \text{GetOperationType}(\text{op})$
    \item \textbf{Select Partitioning Ratio:}
    \begin{align}
    \rho = \begin{cases}
    \rho_{\text{matmul}} & \text{if type} = \text{MatMul} \\
    \rho_{\text{conv}} & \text{if type} = \text{Conv} \\
    \rho_{\text{general}} & \text{otherwise}
    \end{cases} \label{eq:partitioning_ratio}
    \end{align}
    \item \textbf{Calculate Target Cores:} $N_{\text{cores}} = \lceil \rho \times N_{\text{total}} \rceil$
    \item \textbf{Communication-Graph-Aware Placement:} For each operator, compute a placement score per TCC that jointly weighs current load (compute, DMEM, WMEM utilization), NoC hop distance to producer TCCs, workload imbalance penalty, and mesh centrality. Select the TCC with the lowest composite score. This replaces naive round-robin with a placement that minimizes NoC traffic while maintaining load balance.
    \item \textbf{Split Workload:} $\text{workload}_i = \frac{\text{op.workload}}{N_{\text{cores}}}$ for each selected core $i$
\end{enumerate}

The partitioning ratios $\rho_{\text{matmul}}$, $\rho_{\text{conv}}$, and $\rho_{\text{general}}$ are determined by the RL state:
\begin{align}
\rho_{\text{matmul}} &= \text{clip}(\rho_{\text{base}} + \Delta_{\text{matmul}}, 0, 1) \label{eq:matmul_partitioning} \\
\rho_{\text{conv}} &= \text{clip}(\rho_{\text{base}} + \Delta_{\text{conv}}, 0, 1) \label{eq:conv_partitioning} \\
\rho_{\text{general}} &= \text{clip}(\rho_{\text{base}} + \Delta_{\text{general}}, 0, 1) \label{eq:general_partitioning}
\end{align}

where $\rho_{\text{base}} = 0.3$ (default) and $\Delta$ are action deltas from the RL policy. This enables fine-grained load balancing beyond simple node-level assignment.

\subsection{Memory Hierarchy Model}
\label{sec:memory_model}

Each TCC (Tile Compute Cluster) in the mesh has three memory tiers: weight memory (WMEM), data memory (DMEM), and instruction memory (IMEM). The compiler allocates these per-tile based on the operator graph requirements and RL-selected parameters.

\textbf{WMEM capacity constraint:} The total model weight footprint $W_{\text{total}}$ must be distributed across all active tiles:
\begin{align}
\sum_{i=1}^{N_{\text{cores}}} \text{WMEM}_i \geq W_{\text{total}},
\label{eq:wmem_constraint}
\end{align}
where $\text{WMEM}_i$ is the weight memory allocated to tile $i$. For \ModelName{} at \ModelPrecision{}, $W_{\text{total}} = \TotalWeightsMetric{}$.

\textbf{DMEM partitioning:} Data memory ($D_i$ for tile $i$) is split into input, output, and scratch buffers:
\begin{equation}\label{eq:dmem_partition}
D_i = D_i^{\text{in}} + D_i^{\text{out}} + D_i^{\text{scratch}},
\end{equation}
where the allocation fractions are controlled by RL actions (Memory/Load Partition group in Table~\ref{tab:action_space}).

\textbf{Memory bandwidth utilization:} The effective bandwidth per tile depends on access pattern and memory tier:
\begin{equation}\label{eq:bw_utilization}
\text{BW}_{\text{eff},i} = \min\!\!\left(\text{BW}_{\text{pk},i},\; \frac{V_i}{C_i \cdot T_{\text{clk}}}\right)\!,
\end{equation}
where $V_i$ is data volume, $C_i$ is cycle count, and $T_{\text{clk}} = 1/f_{\text{node}}$ is the clock period.

\textbf{Memory pressure metric:} The compiler computes a tile-level memory pressure score that enters the state vector:
\begin{equation}\label{eq:mem_pressure}
\mathcal{P}_i = \frac{W_i^{\text{used}}}{W_i^{\text{alloc}}} + \lambda_d \cdot \frac{D_i^{\text{used}}}{D_i^{\text{alloc}}},
\end{equation}
where $W_i$ and $D_i$ denote WMEM and DMEM for tile $i$, and $\lambda_d = 0.5$ weights data memory pressure relative to weight memory.

\subsection{Network-on-Chip (NoC) Model}
\label{sec:noc_model}

The 2D mesh interconnect carries data between tiles during operator execution. The NoC bandwidth is parameterized by flit width (DFLIT\_WIDTH) which the RL agent selects per-tile.

\textbf{Bisection bandwidth:} For an $M \times N$ mesh, the bisection bandwidth determines the aggregate cross-mesh data rate:
\begin{equation}\label{eq:bisection_bw}
\text{BW}_{\text{bisect}} = \min(M, N) \cdot W_{\text{DFLIT}} \cdot f_{\text{node}},
\end{equation}
where $W_{\text{DFLIT}}$ is the flit width and $f_{\text{node}}$ is the clock frequency.

\textbf{Hop count model:} The average number of hops between two tiles in the mesh determines communication latency:
\begin{align}
\bar{h} = \frac{M + N}{3},\qquad
L_{\text{NoC}} = \bar{h} \times L_{\text{hop}} + L_{\text{setup}},
\label{eq:hop_count}
\end{align}
where $L_{\text{hop}}$ is the per-hop latency and $L_{\text{setup}}$ includes routing header overhead.

\textbf{Communication-to-computation ratio:} This ratio guides the RL agent's mesh sizing decisions:
\begin{align}
\rho_{\text{comm}} = \frac{\sum_{\text{edges}} \text{TensorSize}(e)}{\sum_{\text{ops}} \text{FLOPs}(\text{op})}.
\label{eq:comm_comp_ratio}
\end{align}
A high $\rho_{\text{comm}}$ favors smaller meshes (fewer hops), while compute-dominated workloads benefit from larger meshes with more parallelism.

\subsection{Throughput Model}
\label{sec:throughput_model}

The inference throughput (tokens/s) is bounded by the slowest of three ceilings:

\textbf{Compute ceiling:}
\begin{align}
\text{Tok/s}_{\text{comp}} = \frac{\sum_{i=1}^{N} M_i \cdot 2 \cdot f \cdot \eta_{\parallel} \cdot \alpha_{\text{spec}}}{\text{FLOPs}_{\text{per\_token}}},
\label{eq:compute_ceiling}
\end{align}
where $M_i = \min(\text{TM}_{\text{FP16}},\, \text{VLEN}_i / 16)$ is the effective tensor multiplier count for TCC $i$ (capped by datapath width), $f$ is clock frequency, $\eta_{\parallel}$ is parallel efficiency (Section~\ref{sec:noc_model}), and $\alpha_{\text{spec}}$ is speculative decoding acceleration (1.0--2.0$\times$). $\text{FLOPs}_{\text{per\_token}} = 2 \times P_{\text{total}} \times \phi_{\text{decode}}$ where $P_{\text{total}}$ is total parameters and $\phi_{\text{decode}}$ is the decode-active FLOP fraction ($\approx$0.97 for GQA models).

\textbf{Memory ceiling:}
\begin{align}
\text{Tok/s}_{\text{memory}} = \frac{\sum_i \text{BW}_{\text{eff},i}}{\text{Bytes}_{\text{per\_token}}},
\label{eq:memory_ceiling}
\end{align}
where $\text{Bytes}_{\text{per\_token}}$ accounts for weight reads, KV-cache updates, and activation transfers.

\textbf{NoC ceiling:}
\begin{align}
\text{Tok/s}_{\text{NoC}} = \frac{\text{BW}_{\text{bisect}}}{\text{CrossTileBytes}_{\text{per\_token}}}.
\label{eq:noc_ceiling}
\end{align}

The realized throughput is determined by the binding constraint:
\begin{equation}\label{eq:throughput_bound}
\text{Tok/s} = \min\bigl(T_{\text{comp}},\; T_{\text{mem}},\; T_{\text{NoC}}\bigr).
\end{equation}

where $T_{\text{comp}}$, $T_{\text{mem}}$, and $T_{\text{NoC}}$ are the compute, memory, and NoC ceilings from Eqs.~\ref{eq:compute_ceiling}--\ref{eq:noc_ceiling}. For the Llama 3.1 8B workload, the compute ceiling is the active limiter at all process nodes, as the large mesh sizes and heterogeneous per-TCC VLEN/FETCH saturate compute before memory bandwidth becomes binding.

\subsection{KV-Cache Management and Compaction}
\label{sec:kv_compaction}

Autoregressive decoding in transformer models requires a key-value (KV) cache that grows linearly with sequence length. For \ModelName{} with grouped-query attention (GQA, 8 KV heads), the KV-cache footprint per token is computed at FP16 element width:
\begin{equation}\label{eq:kv_per_token}
\text{KV}_{\text{b/t}} = 2 \cdot n_L \cdot n_{\text{kv}} \cdot d_h \cdot 2,
\end{equation}
where the leading 2 accounts for key and value tensors, $n_L=32$ layers, $n_{\text{kv}}=8$ KV heads, $d_h=128$ head dimension, and the trailing 2 is bytes per FP16 element. This yields $\text{KV}_{\text{b/t}} = 128\,\text{KB}$ per token.

For a sequence length of $L$ tokens, the total KV-cache footprint is:
\begin{align}
\text{KV}_{\text{total}}(L) = L \times \text{KV}_{\text{bytes/tok}}.
\label{eq:kv_total}
\end{align}
At $L = 2048$ (our evaluation setting), $\text{KV}_{\text{total}} = 256\,\text{MB}$, which must be distributed across DMEM allocations on active tiles via Eq.~\ref{eq:dmem_partition}.

\textbf{KV-cache pressure on DMEM.} The KV cache competes with activation scratch space for DMEM capacity. The compiler's DMEM partitioning (controlled by RL actions) must balance:
\begin{align}
\text{DMEM}_i^{\text{in}} &\geq \frac{\text{KV}_{\text{total}}(L)}{N_{\text{active}}} + \text{ActInput}_i, \label{eq:kv_dmem_in} \\
\text{DMEM}_i^{\text{scratch}} &\geq \text{IntermediateBuffer}_i, \label{eq:kv_dmem_scratch}
\end{align}
where $N_{\text{active}}$ is the number of tiles hosting KV-cache slices. If DMEM is undersized, the compiler must spill KV-cache entries to WMEM, increasing latency through the slower memory tier.

\textbf{KV-cache compaction strategies.} To alleviate memory pressure at long sequence lengths, the compiler supports three compaction modes that reduce $\text{KV}_{\text{total}}$:

\textit{(1) Quantized KV cache~\cite{dettmers2022llmint8,xiao2023smoothquant}.} Keys and values are stored in reduced precision (INT8 or INT4) with per-head scale factors:
\begin{align}
\hat{K}_i = \text{round}\!\left(\frac{K_i}{s_K}\right),\quad
\hat{V}_i = \text{round}\!\left(\frac{V_i}{s_V}\right),
\label{eq:kv_quantize}
\end{align}
where $s_K, s_V$ are per-head quantization scales. INT8 quantization halves the KV footprint to $64\,\text{KB}$/token; INT4 reduces it to $32\,\text{KB}$/token.

\textit{(2) Sliding-window eviction.} For layers where full-context attention is not required, a sliding window of size $W$ retains only the most recent tokens:
\begin{align}
\text{KV}_{\text{window}}^{(\ell)} = \min(L, W^{(\ell)}) \times \text{KV}_{\text{bytes/tok}}^{(\ell)},
\label{eq:kv_window}
\end{align}
where $W^{(\ell)}$ can be set per-layer. This is compatible with Llama's rotary position encoding (RoPE), which provides relative position information.

\textit{(3) Paged KV allocation~\cite{kwon2023pagedattention}.} Instead of contiguous KV buffers, the compiler can allocate KV cache in fixed-size pages across tiles:
\begin{align}
N_{\text{pages}} = \left\lceil \frac{\text{KV}_{\text{total}}(L)}{P_{\text{size}}}\right\rceil,
\label{eq:kv_pages}
\end{align}
where $P_{\text{size}}$ is the page size. Paged allocation reduces internal fragmentation when tiles have heterogeneous DMEM capacities, as allocated by the RL agent.

\textbf{Compaction factor.} Combining quantization and windowing, the effective compaction factor is:
\begin{align}
\kappa = \frac{b_{\text{orig}}}{b_{\text{quant}}} \times \frac{L}{\bar{W}},
\label{eq:compaction_factor}
\end{align}
where $b_{\text{orig}}=16$ (FP16), $b_{\text{quant}} \in \{8, 4\}$, and $\bar{W}$ is the mean effective window size across layers. For INT8 quantization with a 1024-token window at $L=2048$, $\kappa = 4\times$, reducing the KV footprint from 256\,MB to 64\,MB.

\textbf{Impact on throughput model.} KV compaction reduces the memory traffic in Eq.~\ref{eq:memory_ceiling}:
\begin{equation}\label{eq:kv_throughput_impact}
B_{\text{tok}}' = B_{\text{tok}} - \!\left(1 - \tfrac{1}{\kappa}\right) \cdot \text{KV}_{\text{b/t}},
\end{equation}
which relaxes the memory ceiling and can shift the binding constraint toward the compute or NoC ceiling. The RL reward function (Eq.~\ref{eq:reward_function}) captures this indirectly through the performance component $P_{\text{norm}}$, as compacted KV caches increase realized throughput.

\subsection{Reward Function}
\label{sec:reward_function}

The reward function balances PPA metrics with adaptive normalization and constraint penalties. The complete reward formulation is:
\begin{align}
R(s, a) ={}& \alpha \times P_{\text{norm}} - \beta \times P_{\text{power}} \nonumber \\
& - \gamma \times A_{\text{norm}} + B_{\text{feasible}} \nonumber \\
& - P_{\text{violation}} - P_{\text{memory}} - P_{\text{hazard}} \label{eq:reward_function}
\end{align}

where each component is defined as:

\textbf{Performance Component:}
\begin{align}
P_{\text{norm}} = \frac{\text{Perf} - \text{Perf}_{\min}}{\text{Perf}_{\max} - \text{Perf}_{\min}} \label{eq:perf_norm}
\end{align}

\textbf{Power Component:}
\begin{align}
P_{\text{power}} = \frac{\text{Power} - \text{Power}_{\min}}{\text{Power}_{\max} - \text{Power}_{\min}} \label{eq:power_norm}
\end{align}

\textbf{Area Component:}
\begin{align}
A_{\text{norm}} = \frac{\text{Area} - \text{Area}_{\min}}{\text{Area}_{\max} - \text{Area}_{\min}} \label{eq:area_norm}
\end{align}

\textbf{Feasibility Bonus:}
\begin{equation}\label{eq:feasibility_bonus}
B_{\text{feasible}} = \begin{cases}
s_{\text{mag}} \cdot (1 + m_{\text{pwr}}) & \text{if feasible} \\
0 & \text{otherwise}
\end{cases}
\end{equation}

\textbf{Violation Penalties:}
\begin{equation}\label{eq:violation_penalty}
P_{\text{violation}} = \begin{cases}
s_{\text{mag}} \cdot (1 + v) \cdot v^2 & \text{if } P > P_{\text{budget}} \\
0 & \text{otherwise}
\end{cases}
\end{equation}

\begin{equation}\label{eq:memory_penalty}
P_{\text{mem}} = \lambda_{\text{mem}} \cdot \max\!\bigl(0,\, M_{\text{used}} - M_{\text{budget}}\bigr)
\end{equation}
\begin{align}
P_{\text{hazard}} &= \lambda_{\text{hazard}} \times \text{TotalHazardScore} \label{eq:hazard_penalty}
\end{align}

where $s_{\text{mag}}$ is the score magnitude, $m_{\text{pwr}} = (P_{\text{budget}} - P)/P_{\text{budget}}$ is the power margin, $v$ is the constraint violation magnitude, and $M_{\text{used}}, M_{\text{budget}}$ are memory used and budget.

\textbf{Adaptive Weights:}
The weights $\alpha$, $\beta$, and $\gamma$ are derived from constraints:
\begin{align}
\alpha &= \frac{w_{\text{perf}}}{w_{\text{perf}} + w_{\text{power}} + w_{\text{area}}} \label{eq:alpha} \\
\beta &= \frac{w_{\text{power}}}{w_{\text{perf}} + w_{\text{power}} + w_{\text{area}}} \label{eq:beta} \\
\gamma &= \frac{w_{\text{area}}}{w_{\text{perf}} + w_{\text{power}} + w_{\text{area}}} \label{eq:gamma}
\end{align}

where $w_{\text{perf}}$, $w_{\text{power}}$, and $w_{\text{area}}$ are user-specified PPA weights (default: 0.4, 0.4, 0.2).

\textbf{Pareto-based final selection.} During RL exploration, every feasible configuration is inserted into a Pareto archive that maintains the non-dominated frontier (Section~\ref{sec:world_model}). After convergence, the final configuration is selected \emph{from the Pareto frontier} using the same weights $(w_{\text{perf}}, w_{\text{power}}, w_{\text{area}})$ as a scalarized selection criterion applied to frontier-normalized objectives. This ensures the returned design is Pareto-optimal---no other explored configuration improves one PPA metric without degrading another.

Table~\ref{tab:reward_components} summarizes the reward function components and their typical ranges.

\begin{table}[ht]
\centering
\footnotesize
\setlength{\tabcolsep}{3pt}
\renewcommand{\arraystretch}{1.1}
\begin{tabular}{@{}lccp{2.4cm}@{}}
\hline
\textbf{Component} & \textbf{Type} & \textbf{Range} & \textbf{Description} \\
\hline
$P_{\text{norm}}$ & Reward & $[0, 1]$ & Norm.\ perf (higher=better) \\
$P_{\text{power}}$ & Penalty & $[0, 1]$ & Norm.\ power (lower=better) \\
$A_{\text{norm}}$ & Penalty & $[0, 1]$ & Norm.\ area (lower=better) \\
$B_{\text{feas}}$ & Bonus & $[0, 2]$ & Feasibility + power margin \\
$P_{\text{viol}}$ & Penalty & $[0, \infty)$ & Cubic constraint viol. \\
$P_{\text{mem}}$ & Penalty & $[0, \infty)$ & Linear memory overuse \\
$P_{\text{haz}}$ & Penalty & $[0, 1]$ & Data hazard penalty \\
\hline
\textbf{Total} & & $[-5, 3]$ & Combined (typical range) \\
\hline
\end{tabular}
\vspace{1pt}
\caption{Reward function components and their characteristics}
\label{tab:reward_components}
\end{table}

Normalization ranges are derived from process node characteristics and constraints, ensuring fair comparison across different technology nodes.

\subsection{Policy Optimization: SAC with Prioritized Replay}
\label{sec:policy_gradient}

The optimizer is Soft Actor-Critic (SAC)~\cite{haarnoja2018soft} with twin Q-networks, auto-tuned entropy, and prioritized experience replay (PER). Table~\ref{tab:sac_config} lists all hyperparameters.

\textbf{Actor-critic architecture.} The actor $\pi_\theta$ and twin critics $Q_{\phi_1}, Q_{\phi_2}$ each use 2-layer MLPs with GELU activation:
\begin{itemize}[nosep]
\item \textbf{Actor:} $[52 \to 256 \to 256 \to 60]$ (30 means + 30 log-stds)
\item \textbf{Critics:} $[82 \to 256 \to 256 \to 1]$ (state-action $\to$ Q-value)
\end{itemize}
Actions are sampled via the reparameterization trick with tanh squashing: $a = \tanh(\mu + \sigma \odot \epsilon)$, $\epsilon \sim \mathcal{N}(0, I)$.

\textbf{Entropy auto-tuning.} The entropy coefficient $\alpha$ is learned with target entropy $\mathcal{H}_{\text{target}} = -d_a = -30$:
\begin{align}
\mathcal{L}_\alpha = -\alpha\,\mathbb{E}[\log \pi(a|s) + \mathcal{H}_{\text{target}}], \label{eq:alpha_loss}
\end{align}
with gradient clipping $\in [-1, 1]$ and $\log\alpha$ bounded to $[-10, 10]$.

\textbf{Critic update.} Twin Q-networks are trained on Bellman residuals with clipped double-Q targets:
\begin{align}
y_t &= r_t + \gamma\bigl[\min_{i} Q_{\bar{\phi}_i}(s', a') - \alpha \log \pi(a'|s')\bigr], \label{eq:q_target} \\
\mathcal{L}_Q &= \mathbb{E}_{(s,a,r,s') \sim \mathcal{D}}\bigl[(Q_{\phi_i}(s,a) - y_t)^2\bigr], \label{eq:q_loss}
\end{align}
where $\bar{\phi}_i$ are soft-updated target networks with $\tau = 0.005$.

\textbf{Prioritized replay buffer.} Transitions are stored in a 100K-capacity buffer with stochastic prioritized sampling (priority exponent $\alpha_{\text{PER}} = 0.6$, importance sampling $\beta = 0.4 \to 1.0$ annealed at $+0.001$ per sample). Priorities are set from TD-error: $p_i = (|\delta_i| + 10^{-6})^{0.6}$.

\begin{table}[ht]
\centering
\scriptsize
\setlength{\tabcolsep}{2pt}
\begin{tabular}{@{}lp{1.8cm}r@{}}
\hline
& \textbf{Parameter} & \textbf{Value} \\
\hline
\multirow{4}{*}{\rotatebox{90}{\tiny SAC}}
& Hidden layers & [256, 256] \\
& LR ($\pi$, $Q$, $\alpha$) & $3\!\times\!10^{-4}$ \\
& Batch / $\tau$ & 256 / 0.005 \\
& Warmup & 1,000 \\
\hline
\multirow{2}{*}{\rotatebox{90}{\tiny PER}}
& Buffer cap. & 100K \\
& $\alpha$ / $\beta$ & 0.6 / 0.4$\to$1 \\
\hline
\multirow{3}{*}{\rotatebox{90}{\tiny MPC}}
& $K$ / $H$ & 64 / 5 \\
& Noise $\sigma$ & 0.3 \\
& Blend & 70/30\% \\
\hline
\rotatebox{90}{\tiny WM}
& Arch & 82-128-64-52 \\
\hline
\end{tabular}
\caption{SAC, PER, MPC, and world model config.}
\label{tab:sac_config}
\end{table}

Algorithm~\ref{alg:rl_loop} formalizes the complete optimization loop.

\refstepcounter{algorithm}
\begin{center}
\fbox{\parbox{0.92\columnwidth}{
\footnotesize
\textbf{Algorithm \thealgorithm:} Unified RL-Based Hardware-Aware Compilation\\[-3pt]
\rule{\linewidth}{0.4pt}\\[2pt]
\textbf{Input:} Model graph $G$, nodes $\mathcal{N}$, PPA weights $(w_p,w_w,w_a)$, budget $T$, schedule $\epsilon_0 \!\to\! \epsilon_{\min}$\\
\textbf{Output:} Best configuration per node $\{c^*_n\}_{n \in \mathcal{N}}$\\[2pt]
1: Initialize policy $\pi_\theta$, baseline $b \leftarrow 0$, $\epsilon \leftarrow \epsilon_0$\\
2: \textbf{for each} node $n \in \mathcal{N}$ \textbf{do}\\
3: \quad Load constraints $\mathcal{C}_n$; init mesh $m \leftarrow m_0(n)$, $s^* \!\leftarrow\! \infty$\\
4: \quad \textbf{for} $t = 1$ to $T_n$ \textbf{do}\\
5: \quad\quad $\mathbf{s}_t \leftarrow \text{Encode}(G, m, \mathcal{C}_n)$\\
6: \quad\quad \textbf{if} $\text{rand}() < \epsilon$ \textbf{then} $\mathbf{a}_t \sim \text{Uniform}$
          \textbf{else} $\mathbf{a}_t \sim \pi_\theta(\cdot \mid \mathbf{s}_t)$\\
7: \quad\quad Project: $\mathbf{a}_t' \leftarrow \Pi_{\mathcal{C}_n}(\mathbf{a}_t)$\\
8: \quad\quad Apply mesh deltas + per-TCC updates from $\mathbf{a}_t'$\\
9: \quad\quad Partition operators across TCCs (Sec.~3.5)\\
10:\quad\quad $r_t \leftarrow R(\mathbf{s}_t, \mathbf{a}_t')$ \hfill\textit{// Eq.~\ref{eq:reward_function}}\\
11:\quad\quad Store $(s_t, a_t, r_t, s_{t+1})$ in PER buffer $\mathcal{D}$\\
12:\quad\quad Sample mini-batch (256) from $\mathcal{D}$; update $Q_{\phi_{1,2}}, \pi_\theta, \alpha$\\
13:\quad\quad Train world model $f_\omega$ on $\Delta s$ from batch\\
14:\quad\quad \textbf{if} $f_\omega$ trained \textbf{and} $\epsilon < 0.15$: MPC-refine $\mathbf{a}_t$\\
15:\quad\quad $\epsilon \leftarrow \max(\epsilon_{\min},\, \epsilon \times d_\epsilon)$\\
16:\quad\quad \textbf{if} PPA $< s^*$ and feasible \textbf{then} $s^* \!\leftarrow$ PPA; $c^*_n \!\leftarrow$ config\\
15:\quad \textbf{end for}\\
16:\quad Emit RTL artifacts for $c^*_n$\\
17: \textbf{end for}\\
18: \textbf{return} $\{c^*_n\}_{n \in \mathcal{N}}$
}}
\label{alg:rl_loop}
\end{center}

Table~\ref{tab:rl_hyperparameters} summarizes the key RL hyperparameters used in our design methodology.

\begin{table*}[!htbp]
\centering
\small
\setlength{\tabcolsep}{3pt}
\renewcommand{\arraystretch}{1.12}
\begin{tabular}{lllp{4cm}}
\hline
\textbf{Hyperparameter} & \textbf{Value} & \textbf{Component} & \textbf{Description} \\
\hline
Actor LR / Critic LR / $\alpha$ LR & $3 \times 10^{-4}$ & SAC & All three learning rates \\
Discount Factor ($\gamma$) & 0.99 & SAC & Future reward discount \\
Soft Target Update ($\tau$) & 0.005 & SAC & Polyak averaging for target Q-networks \\
Initial Entropy ($\alpha_0$) & 0.2 & SAC & Auto-tuned via Eq.~\ref{eq:alpha_loss} \\
Target Entropy & $-30$ & SAC & $-d_a$ (negative action dim) \\
Mini-batch Size & 256 & SAC & Sampled from PER buffer \\
Warmup Steps & 1,000 & SAC & Collect experience before training \\
Replay Buffer Capacity & 100,000 & PER & Max stored transitions \\
Priority Exponent ($\alpha_{\text{PER}}$) & 0.6 & PER & Prioritization strength \\
IS Exponent ($\beta$) & 0.4 $\to$ 1.0 & PER & Annealed at $+0.001$ per sample \\
Exploration ($\epsilon$) & 0.5 $\to$ 0.1 & $\epsilon$-greedy & Auto-derived $d$; slowed $d'$ when stuck \\
State Dimension & 52 & All & Optimized feature subset \\
Action Dimension & 30 (80 policy) & All & 4 discrete (20 one-hot) + 30 cont.\ ($\times 2$ for $\mu, \log\sigma$) \\
Actor / Critic Hidden & $2 \times 256$ & SAC & Two-layer MLP with GELU \\
World Model Hidden & $[128, 64]$ & MPC & Residual $\Delta s$ prediction \\
MPC Candidates / Horizon & 64 / 5 & MPC & Planning with $\gamma = 0.99$ \\
MPC Blend Ratio & 70\% / 30\% & MPC & MPC vs SAC for TCC params \\
\hline
\end{tabular}
\vspace{1pt}
\caption{Complete RL system hyperparameters: SAC optimizer, prioritized replay, and MPC planner.}
\label{tab:rl_hyperparameters}
\end{table*}

\subsection{Per-TCC Parameter Constraints}

Table~\ref{tab:tcc_constraints} lists the per-TCC parameters controlled by the RL agent and their valid ranges. These constraints are node-dependent: smaller process nodes permit higher frequencies and tighter voltage margins, which expand the feasible region for memory and compute parameters.

\begin{table*}[!htbp]
\centering
\small
\setlength{\tabcolsep}{8pt}
\renewcommand{\arraystretch}{1.15}
\begin{tabular}{llll}
\hline
\textbf{Parameter} & \textbf{Min} & \textbf{Max} & \textbf{Notes} \\
\hline
FETCH\_SIZE & 1 & 16 & Instruction fetch width (per-TCC) \\
STANUM & 1 & 32 & Reservation stations \\
VLEN (bits) & 128 & 2048 & Vector register length (per-TCC) \\
DMEM\_SIZE\_KB & 16 & 512 & Data memory per tile \\
WMEM\_SIZE\_KB & 256 & adaptive & Weight ROM (model-dependent) \\
IMEM\_SIZE\_KB & 1 & 128 & Instruction memory per tile \\
DFLIT\_WIDTH & 64 & 8192 & NoC flit width (bits, chip-level) \\
XR\_WP & 1 & 16 & Scalar register write ports \\
VR\_WP & 1 & 16 & Vector register write ports \\
XDPNUM & 1 & 16 & Scalar dispatch ports \\
VDPNUM & 1 & 16 & Vector dispatch ports \\
\hline
\end{tabular}
\vspace{1pt}
\caption{Per-TCC parameter ranges (11 parameters). Bounds are architectural limits; the RL agent selects continuous values within these bounds, which are then quantized to hardware-supported discrete values.}
\label{tab:tcc_constraints}
\end{table*}

\subsection{Reward Sensitivity Analysis}
\label{sec:sensitivity}

The PPA reward weights $(w_{\text{perf}}, w_{\text{power}}, w_{\text{area}})$ directly influence the selected configuration. We characterize this sensitivity by analyzing the gradient of the reward function with respect to each weight:
\begin{align}
\frac{\partial R}{\partial w_{\text{perf}}} &= \frac{P_{\text{norm}} \cdot (w_{\text{power}} + w_{\text{area}})}{(w_{\text{perf}} + w_{\text{power}} + w_{\text{area}})^2}, \label{eq:sensitivity_perf}
\end{align}
with analogous expressions for $w_{\text{power}}$ and $w_{\text{area}}$. For the performance-priority mode used in this paper $(w_{\text{perf}}=0.4, w_{\text{power}}=0.4, w_{\text{area}}=0.2)$, the resulting normalized weights are $\alpha=0.4$, $\beta=0.4$, $\gamma=0.2$, which balances throughput against power while treating area as a secondary objective. Shifting to an area-priority configuration $(0.2, 0.2, 0.6)$ would favor compact meshes at the cost of throughput.

\subsection{Complexity and Scalability}

The per-episode cost is dominated by PPA evaluation (codegen + simulation), which runs in $O(N_{\text{ops}} \times N_{\text{cores}})$ for operator partitioning and $O(N_{\text{cores}})$ for per-TCC configuration. The policy network forward pass is $O(|\mathbf{s}| \times H + H \times |\mathbf{a}|)$ where $H=128$, negligible relative to PPA evaluation. The total search cost for one node is:
\begin{align}
C_{\text{total}}(n) = T_n \times \left(C_{\text{policy}} + C_{\text{ppa}}(n)\right),
\end{align}
where $T_n$ is the episode budget and $C_{\text{ppa}}(n)$ is the node-dependent evaluation cost. Across all nodes, the compiler runs sequentially:
\begin{align}
C_{\text{all}} = \sum_{n \in \mathcal{N}} C_{\text{total}}(n).
\end{align}
The surrogate model (Section~\ref{sec:sac_surrogate}) can amortize $C_{\text{ppa}}$ by pre-filtering candidate actions, reducing the number of full evaluations per episode. For meshes larger than 50$\times$50, hierarchical decomposition (block-level RL followed by intra-block tuning) offers a path to sub-linear scaling with mesh size.

\subsection{RL Network, MoE Policy, SAC, and Surrogate Modeling (Llama Example)}
\label{sec:sac_surrogate}

The production compilation flow uses Soft Actor-Critic (SAC) with entropy-regularized exploration (Section~\ref{sec:policy_gradient}) and Mixture-of-Experts (MoE) gating for the policy network. The same state/action interface also supports REINFORCE and PPO for simpler workloads.

\textbf{Policy network (actor):}
\begin{equation}
\pi_\theta(a \mid s) = \pi_\theta(a_d \mid s)\,\pi_\theta(a_c \mid s),
\end{equation}
where $a_d$ are discrete mesh actions (e.g., width/height deltas) and $a_c$ are continuous per-core controls (e.g., memory and fetch tuning).
\begin{align}
\mathcal{L}_{\text{PG}}(\theta)= -\mathbb{E}_{(s_t,a_t)\sim \mathcal{D}}
\!\left[\log \pi_\theta(a_t\mid s_t)\,\hat{A}_t\right],
\end{align}
\begin{align}
\hat{A}_t &= \sum_{\ell=0}^{T-t-1}(\gamma\lambda)^\ell \delta_{t+\ell}, \nonumber \\
\delta_t &= r_t+\gamma V_\nu(s_{t+1})-V_\nu(s_t).
\end{align}
This generalized advantage formulation reduces to REINFORCE when $\lambda=1$ (no value baseline) and provides lower-variance gradients when the SAC critic is available. In production, SAC uses this with its learned Q-functions as the advantage estimator.

\textbf{MoE policy head:}
\begin{align}
g_k(s) &= \frac{\exp(u_k^\top s)}{\sum_{j=1}^{K}\exp(u_j^\top s)}, \nonumber \\
\pi_\theta(a\mid s) &= \sum_{k=1}^{K} g_k(s)\,\pi_{\theta_k}(a\mid s).
\end{align}
The gating network $g_k(s)$ routes each state to expert policies $\pi_{\theta_k}$, which is useful when different Llama operator regimes (attention, MLP, memory-heavy phases) require distinct action preferences.
\begin{align}
\mathcal{L}_{\text{MoE-balance}} &=
\lambda_{\text{lb}}\,K\sum_{k=1}^{K}\bar{g}_k^2, \nonumber \\
\bar{g}_k &= \frac{1}{B}\sum_{b=1}^{B}g_k(s_b),
\end{align}
which penalizes expert collapse and improves routing diversity across compiler states.

\textbf{Critic network:}
\begin{align}
Q_\phi(s,a) \approx \mathbb{E}\!\left[\sum_{t=0}^{T-1}\gamma^t r_t \,\middle|\, s_0=s,a_0=a\right].
\end{align}
With actor-critic, the actor is updated against critic estimates, and the critic is updated by Bellman targets.
\begin{align}
\mathcal{L}_{Q}(\phi_i)=
\mathbb{E}_{(s_t,a_t,r_t,s_{t+1})\sim\mathcal{D}}
\!\left[\!\left(Q_{\phi_i}(s_t,a_t)-y_t\right)^2\right].
\end{align}

\textbf{SAC objective:}
\begin{align}
J_\pi(\theta) ={}& \mathbb{E}_{s_t \sim \mathcal{D},\, a_t \sim \pi_\theta}
\Big[\alpha_{\text{ent}} \log \pi_\theta(a_t \mid s_t) \nonumber \\
&\quad - Q_\phi(s_t,a_t)\Big],
\end{align}
\begin{align}
y_t ={}& r_t + \gamma\,\mathbb{E}_{a_{t+1}\sim \pi_\theta}\!\Big[
\min_{i\in\{1,2\}} Q_{\bar{\phi}_i}(s_{t+1},a_{t+1}) \nonumber \\
&\qquad - \alpha_{\text{ent}}\log\pi_\theta(a_{t+1}\mid s_{t+1})
\Big].
\end{align}
\begin{align}
\mathcal{L}_{\alpha}=
\mathbb{E}_{a_t\sim\pi_\theta}\!\left[
-\alpha_{\text{ent}}\left(\log\pi_\theta(a_t\mid s_t)+\mathcal{H}_{\text{target}}\right)\right],
\end{align}
where $\alpha_{\text{ent}}$ is learned to maintain target entropy under changing node constraints.
This entropy-regularized objective is robust in broad hardware design spaces and is compatible with our mixed discrete/continuous action heads.

\textbf{Surrogate model for PPA:}
\begin{align}
\hat{\mathbf{m}}_\psi(s,a)={}&\left[\hat{P}_{\text{power}},\hat{P}_{\text{perf}},\hat{P}_{\text{area}}\right], \nonumber \\
\hat{r}(s,a)={}&f_{\text{ppa}}\!\left(\hat{\mathbf{m}}_\psi(s,a)\right).
\end{align}

Each surrogate output head is process-node-dependent. The power prediction $\hat{P}_{\text{power}}$ decomposes into compute logic power and memory read power:
\begin{align}
\hat{P}_{\text{power}}(s,a,n) ={}&
  N_{\text{cores}} \cdot \left[P_{\text{logic}}(a) \cdot \kappa_P(n)\right] \nonumber \\
  &+ W_{\text{total}} \cdot E_{\text{dyn}}(n) \cdot \alpha,
\end{align}
where $\kappa_P(n) = \sqrt{A_{\text{scale}}(n)} \cdot V_{\text{dd}}^2(n)$ is the node-dependent power scaling factor (relative to 28nm), $E_{\text{dyn}}(n)$ is the per-MB dynamic read energy at node $n$, and $\alpha$ is the memory access activity factor---both interpolated from the foundry-calibrated process node table. ROM (weight memory) static leakage is eliminated by sleep transistors inserted on the Vdd rail during backend physical design; only SRAM (activation/instruction memory) retains peripheral leakage.

The clock frequency $f_{\text{clk}}(n)$ is an RL-optimized parameter bounded by each process node's maximum achievable frequency. In high-performance mode the RL agent pins the clock to the node maximum, yielding 1\,GHz at 3nm down to 250\,MHz at 28nm. The performance prediction $\hat{P}_{\text{perf}}$ thus scales with core count and node-dependent clock:
\begin{align}
\hat{P}_{\text{perf}}(s,a,n) ={}& N_{\text{cores}} \cdot \text{VLEN} \nonumber \\
  &\cdot f_{\text{clk}}(n) \cdot \eta_{\text{util}}(s,a),
\end{align}
where $f_{\text{clk}}(n)$ is the clock frequency at node $n$ (e.g.\ 1\,GHz at 3nm, 820\,MHz at 5nm, 250\,MHz at 28nm) and $\eta_{\text{util}}$ captures pipeline utilization efficiency predicted from workload features and memory pressure.

The area prediction $\hat{P}_{\text{area}}$ combines logic and memory area:
\begin{align}
\hat{P}_{\text{area}}(s,a,n) ={}& N_{\text{cores}} \cdot A_{\text{logic}} \cdot A_{\text{scale}}(n) \nonumber \\
  &+ W_{\text{total}} \cdot A_{\text{ROM/MB}}(n) \nonumber \\
  &+ D_{\text{total}} \cdot A_{\text{SRAM/MB}}(n),
\end{align}
where $A_{\text{scale}}(n)$, $A_{\text{ROM/MB}}(n)$, and $A_{\text{SRAM/MB}}(n)$ are interpolated from the process node table.

The surrogate loss and acceptance criterion are:
\begin{align}
\mathcal{L}_{\text{sur}}(\psi)=
\sum_{q} w_q
\left\|m_q-\hat{m}_{q,\psi}(s,a)\right\|_2^2,
\end{align}
where $q \in \{\text{pwr, perf, area}\}$.
\begin{align}
\sigma_\psi^2(s,a) = \tfrac{1}{3}\sum_q \left(m_q-\hat{m}_{q,\psi}(s,a)\right)^2,
\end{align}
\begin{equation}
\mathbf{1}_{\text{accept}} = \mathbf{1}[\sigma_\psi^2<\tau_{\text{sur}}].
\end{equation}
The surrogate provides fast PPA estimates for candidate actions before expensive full evaluation, and uncertainty-gated usage can be enforced by accepting surrogate predictions only when confidence exceeds a threshold. The explicit node dependence in each output head ensures that the surrogate generalizes across the \ProcessNodeRange{} range without retraining.

\textbf{Llama 3.1 8B example:}
for our Llama workload, state $s$ includes model/workload descriptors (operators, tensor-interface pressure), node constraints, and current mesh/per-core configuration; action $a$ proposes mesh and per-core updates; reward follows Eq.~\ref{eq:reward_function}. A constrained action projection is applied before evaluation:
\begin{align}
a_t' &= \Pi_{\mathcal{C}_{\text{node}}}(a_t), \nonumber \\
\mathcal{C}_{\text{node}} &= \{a:\ P(a)\le P_{\max},\ A(a)\le A_{\max}\}.
\end{align}
In this paper's measured run, this loop converges to \BestNode{} with mesh \BestMesh{} and PPA \BestPPA{}, while the same interface can train SAC/actor-critic/MoE variants without changing optimization targets.

\subsection{World Model and Model-Predictive Planning}
\label{sec:world_model}

\textbf{World model.} A 2-layer MLP $f_\omega: \mathbb{R}^{82} \to \mathbb{R}^{52}$ (hidden dims $[128, 64]$, GELU activation) predicts state deltas via residual learning:
\begin{align}
\hat{s}_{t+1} = s_t + f_\omega([s_t; a_t]),
\end{align}
where $[\cdot;\cdot]$ denotes concatenation of the 52-dim state and 30-dim action. The model is trained online from SAC replay transitions with MSE loss on $\Delta s = s_{t+1} - s_t$ at half the critic learning rate. Residual prediction is stable because consecutive design states differ by small perturbations (mesh $\pm 1$, memory $\pm$ one bank).

\textbf{MPC planning.} Once the world model is trained, Model-Predictive Control activates during exploitation ($\epsilon < 0.15$). For each decision point, $K\!=\!64$ candidate action sequences are evaluated over horizon $H\!=\!5$:
\begin{align}
a^{(i)}_0 &= \text{clamp}(\pi_\theta(s_t) + \epsilon_i,\, {-}1, 1), \\
\hat{s}^{(i)}_{k+1} &= \hat{s}^{(i)}_k + f_\omega([\hat{s}^{(i)}_k;\, \pi_\theta(\hat{s}^{(i)}_k)]), \\
G^{(i)} &= \textstyle\sum_{k=0}^{H-1} \gamma^k\, r_{\text{sur}}(\hat{s}^{(i)}_k),
\end{align}
where $\epsilon_i \sim \mathcal{N}(0, 0.3^2)$, $k \geq 1$ uses the policy for future actions, and $r_{\text{sur}} = \hat{P}_{\text{perf}} - 0.3\,\hat{P}_{\text{pwr}} - 0.2\,\hat{P}_{\text{area}}$ is the surrogate PPA reward.
where the surrogate PPA head evaluates each rolled-out state. The action from $\arg\max_i G^{(i)}$ is blended with the SAC policy: $a_{\text{final}} = 0.7\,a_{\text{MPC}} + 0.3\,a_{\text{SAC}}$ for continuous TCC parameters (action indices 20--31: FETCH, STANUM, VLEN, DMEM, WMEM, IMEM, DFLIT, ports). Discrete mesh actions (indices 0--19) remain SAC-only, as MPC's continuous perturbations are ill-suited for discrete topology decisions.

The planning cost is $K \times H = 320$ forward passes through the lightweight world model and surrogate heads---under 1\,ms total, negligible versus the $\sim$10\,ms full PPA evaluation. MPC provides multi-step lookahead that helps the policy navigate correlated parameter interactions (e.g., increasing VLEN while decreasing mesh size) that single-step SAC may explore inefficiently.

\section{Results and Evaluation}

\subsection{Experimental Setup}

We evaluate on:
\begin{itemize}
    \item \textbf{Model:} Llama 3.1 8B Instruct FP16 ONNX
    \item \textbf{Workload:} \TotalWeightsMetric{} weights, \GraphOperators{} graph operators, 291 weight tensors
    \item \textbf{Process Nodes:} 3nm, 5nm, 7nm, 10nm, 14nm, 22nm, 28nm
    \item \textbf{Mode:} RL performance-priority optimization
    \item \textbf{Metrics:} Power (mW), Performance (GOps/s, counting FP16 multiply-accumulate operations), Area (mm\textsuperscript{2}), PPA Score, Tokens/s
\end{itemize}

Table~\ref{tab:experimental_setup} summarizes the experimental configuration.

\begin{table*}[!htbp]
\centering
\footnotesize
\setlength{\tabcolsep}{3pt}
\renewcommand{\arraystretch}{1.1}
\begin{tabular}{lcp{2.8cm}}
\hline
\textbf{Component} & \textbf{Value} & \textbf{Description} \\
\hline
Target Model & \ModelName{} \ModelPrecision{} & Decoder-only transformer model \\
Total Weights & \TotalWeightsMetric{} & Weight tensors mapped to WMEM \\
Operators & \GraphOperators{} & Unified graph operator count \\
Weight Tensors & 291 & Initializer tensors used by codegen \\
Inputs / Outputs & 66 / 65 & Graph interface tensors \\
Process Nodes & 3, 5, 7, 10, 14, 22, 28 nm & Technology nodes evaluated \\
RL Episodes & Up to \MaxEpisodesThreeNm{} & Node-adaptive exploration budget \\
Evaluation Configs & 7 & Process nodes × 1 model \\
PPA Metrics & 5 & Power, Performance, Area, Score, Tokens/s \\
\hline
\end{tabular}
\vspace{1pt}
\caption{Experimental setup summary}
\label{tab:experimental_setup}
\end{table*}
\subsection{RL Training Convergence}

The reinforcement learning optimization demonstrates robust convergence for Llama 3.1 8B in performance-priority mode. Training exhibits three phases: (1) \textit{Initial Exploration} with high reward variance and broad mesh searches, (2) \textit{Learning Phase} with systematic reward improvement, and (3) \textit{Convergence Phase} where policy updates stabilize around node-specific optima.

The adaptive exploration mechanism dynamically adjusts the exploration rate from $\epsilon = 0.5$ to $\epsilon = 0.1$, enabling smooth transition from exploration to exploitation. In the final run, RL search converges within \MaxEpisodesThreeNm{} episodes for the \BestNode{} node, with the globally best configuration selected at \BestNode{} (PPA score \BestPPA{}).

\begin{figure*}[ht]
\centering
\includegraphics[width=\textwidth]{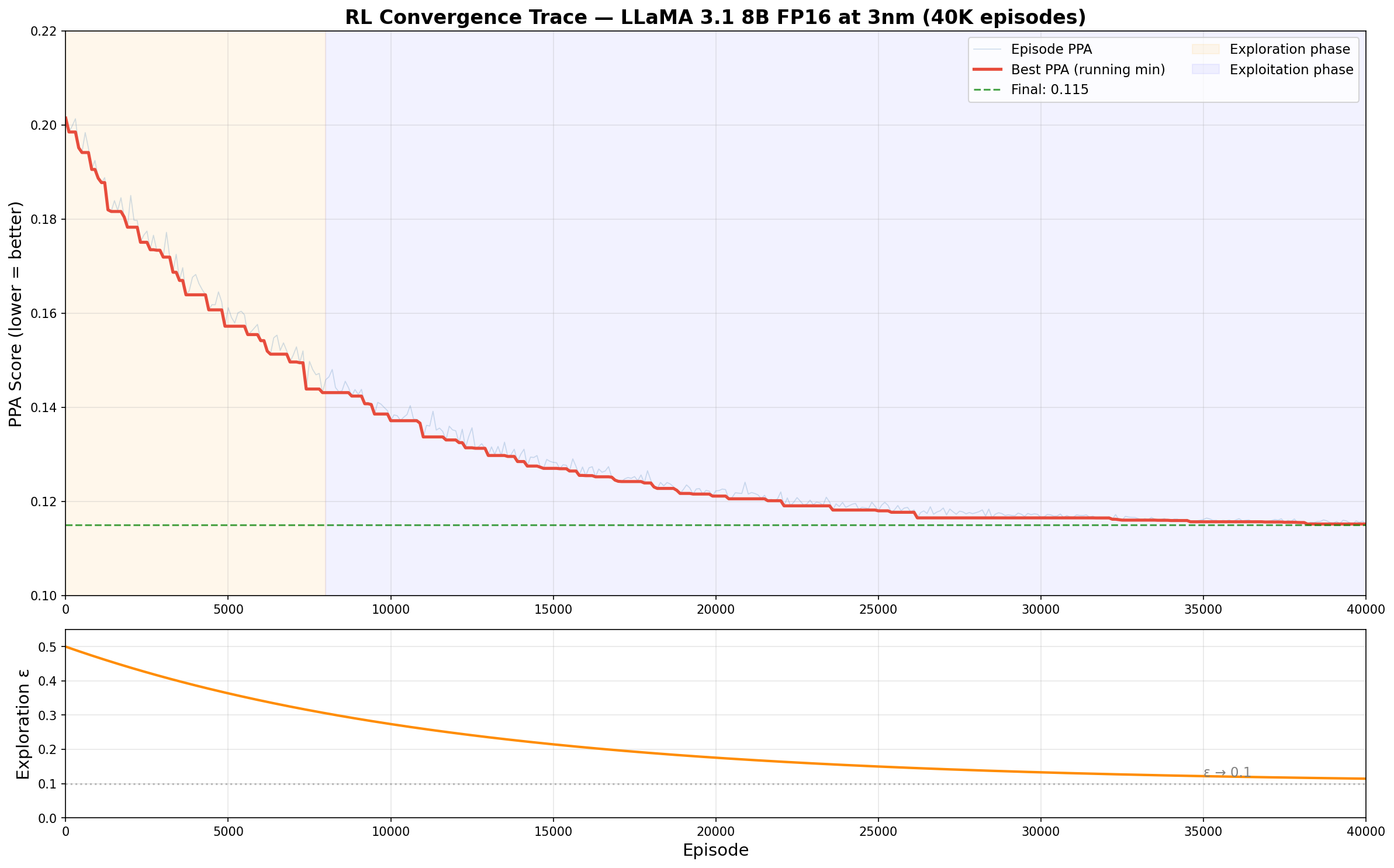}
\caption{RL convergence trace at 3nm: best PPA score vs.\ episode count over $\sim$4.6K episodes, showing exploration-to-exploitation transition.}
\label{fig:rl_convergence}
\end{figure*}

Figure~\ref{fig:rl_convergence} uses parsed compiler-log traces from the latest run and shows exploration saturation (unique configurations) together with policy entropy stabilization.

\subsection{Llama 3.1 8B Model Characteristics}

\ModelName{}~\cite{grattafiori2024llama3} is a decoder-only transformer with 32 layers, grouped-query attention (8 KV heads), and a 128K-token context window. At \ModelPrecision{} precision, the \TotalWeightsMetric{} weight footprint (\TotalParameters{} parameters) presents a strongly memory-dominated optimization problem. The model comprises \GraphOperators{} graph operators and \TotalInstructions{} total instructions. Evaluation uses a \MaxSequenceLength{}-token sequence length in \OptimizationMode{} mode. Table~\ref{tab:llama_characteristics} provides the key compilation statistics.

\begin{table*}[!htbp]
\centering
\footnotesize
\begin{filecontents*}{data/model_characteristics_actual.csv}
characteristic,value,description
Model Family,Llama 3.1 8B FP16,Decoder-only transformer
Parameters,8.03B,Total model parameters
Operators,7489,Unified graph operators
Total Weights,14.96 GB,FP16 weight footprint
Best Mesh,41x42,Best node mesh (3nm)
Best Throughput,29809 tok/s,Batch=3 seq-len=2048
Evaluated Nodes,7,Process nodes: 3 5 7 10 14 22 28 nm
\end{filecontents*}
\pgfplotstabletypeset[
    col sep=comma,
    string type,
    columns={characteristic,value,description},
    columns/characteristic/.style={column name=\textbf{Characteristic}},
    columns/value/.style={column name=\textbf{Value}},
    columns/description/.style={column name=\textbf{Description}},
    every head row/.style={before row=\hline,after row=\hline},
    every last row/.style={after row=\hline}
]{data/model_characteristics_actual.csv}
\vspace{1pt}
\caption{Llama 3.1 8B FP16 model characteristics and run statistics}
\label{tab:llama_characteristics}
\end{table*}

The model presents unique optimization challenges:
\begin{itemize}[leftmargin=*]
    \item \textbf{Large weight footprint:} \TotalWeightsMetric{} weights require aggressive WMEM-aware placement (Eq.~\ref{eq:wmem_constraint}).
    \item \textbf{KV-cache pressure:} 66 input and 65 output tensors increase DMEM demand; compaction strategies (Section~\ref{sec:kv_compaction}) control memory growth.
    \item \textbf{Node-dependent tradeoffs:} Throughput, area, and power shift non-linearly across nodes.
    \item \textbf{Compute-limited throughput:} Ceiling analysis (Eq.~\ref{eq:throughput_bound}) identifies compute as the binding constraint at all nodes.
    \item \textbf{Joint PPA tradeoffs:} Score, tok/s, area, and power vary non-linearly with node (Section~\ref{sec:ppa_correlation}).
\end{itemize}

\subsection{PPA Results Across Process Nodes}

Table~\ref{tab:mesh_scaling} shows the RL outcomes by process node. The observed scaling is empirical and reflects node-specific tradeoffs between mesh size, power, throughput, and area.

Table~\ref{tab:mesh_scaling} provides detailed mesh configurations and their scaling factors relative to the best node (3nm). The mesh sizes decrease monotonically from 1,722 TCCs (3nm) to 132 TCCs (28nm), reflecting both the smaller die-area budget at larger nodes and the RL agent's adaptation to node-specific clock and density constraints.

\begin{table*}[!htbp]
\centering
\footnotesize
\setlength{\tabcolsep}{4pt}
\renewcommand{\arraystretch}{1.1}
\begin{filecontents*}{data/mesh_scaling_actual.csv}
process_node,mesh_size,total_cores,freq_mhz,scaling_factor,power_mw,perf_gops,area_mm2,ppa_score
3nm,41x42,1722,1000,1.00x,51366,466364,648,0.974
5nm,39x39,1521,820,0.88x,57153,338116,929,0.989
7nm,33x34,1122,570,0.65x,46208,173899,1220,0.996
10nm,26x27,702,520,0.41x,25134,99939,1572,1.005
14nm,21x22,462,400,0.27x,14161,51072,1992,1.016
22nm,16x16,256,250,0.15x,7093,18077,2882,1.023
28nm,11x12,132,250,0.08x,3780,9744,3545,1.019
\end{filecontents*}
\pgfplotstabletypeset[
    col sep=comma,
    string type,
    columns={process_node,mesh_size,total_cores,scaling_factor,power_mw,perf_gops,area_mm2,ppa_score},
    columns/process_node/.style={column name=\textbf{Node}},
    columns/mesh_size/.style={column name=\textbf{Mesh}},
    columns/total_cores/.style={column name=\textbf{Cores}},
    columns/freq_mhz/.style={column name=\textbf{Freq (MHz)}},
    columns/scaling_factor/.style={column name=\textbf{Scaling}},
    columns/power_mw/.style={column name=\textbf{Power (mW)}},
    columns/perf_gops/.style={column name=\textbf{Perf (GOps)}},
    columns/area_mm2/.style={column name=\textbf{Area (mm\textsuperscript{2})}},
    columns/ppa_score/.style={column name=\textbf{PPA}},
    every head row/.style={before row=\hline,after row=\hline},
    every last row/.style={after row=\hline}
]{data/mesh_scaling_actual.csv}
\vspace{1pt}
\caption{Per-process-node RL results for Llama 3.1 8B FP16. Scaling factor is core count relative to 3nm (1722 cores).}
\label{tab:mesh_scaling}
\end{table*}

\begin{table*}[!htbp]
\centering
\footnotesize
\begin{filecontents*}{data/ppa_by_node_actual.csv}
process_node,mesh_config,cores,freq_mhz,power_mw,perf_gops,area_mm2,ppa_score,tok_s
3nm,41x42,1722,1000,51366,466364,648,0.974,29809
5nm,39x39,1521,820,57153,338116,929,0.989,21612
7nm,33x34,1122,570,46208,173899,1220,0.996,11115
10nm,26x27,702,520,25134,99939,1572,1.005,6388
14nm,21x22,462,400,14161,51072,1992,1.016,3264
22nm,16x16,256,250,7093,18077,2882,1.023,1155
28nm,11x12,132,250,3780,9744,3545,1.019,623
\end{filecontents*}
\pgfplotstabletypeset[
    col sep=comma,
    string type,
    columns={process_node,mesh_config,cores,freq_mhz,power_mw,perf_gops,area_mm2,ppa_score,tok_s},
    columns/process_node/.style={column name=\textbf{Node}},
    columns/mesh_config/.style={column name=\textbf{Mesh}},
    columns/cores/.style={column name=\textbf{Cores}},
    columns/freq_mhz/.style={column name=\textbf{Freq (MHz)}},
    columns/power_mw/.style={column name=\textbf{Power (mW)}},
    columns/perf_gops/.style={column name=\textbf{Perf (GOps)}},
    columns/area_mm2/.style={column name=\textbf{Area (mm\textsuperscript{2})}},
    columns/ppa_score/.style={column name=\textbf{PPA}},
    columns/tok_s/.style={column name=\textbf{Tok/s}},
    every head row/.style={before row=\hline,after row=\hline},
    every last row/.style={after row=\hline}
]{data/ppa_by_node_actual.csv}
\vspace{1pt}
\caption{Optimized PPA metrics across process nodes for Llama 3.1 8B FP16}
\label{tab:ppa_by_node}
\end{table*}

\begin{figure*}[!htbp]
\centering
\begin{subfigure}[t]{0.48\textwidth}
\centering
\includegraphics[width=\textwidth]{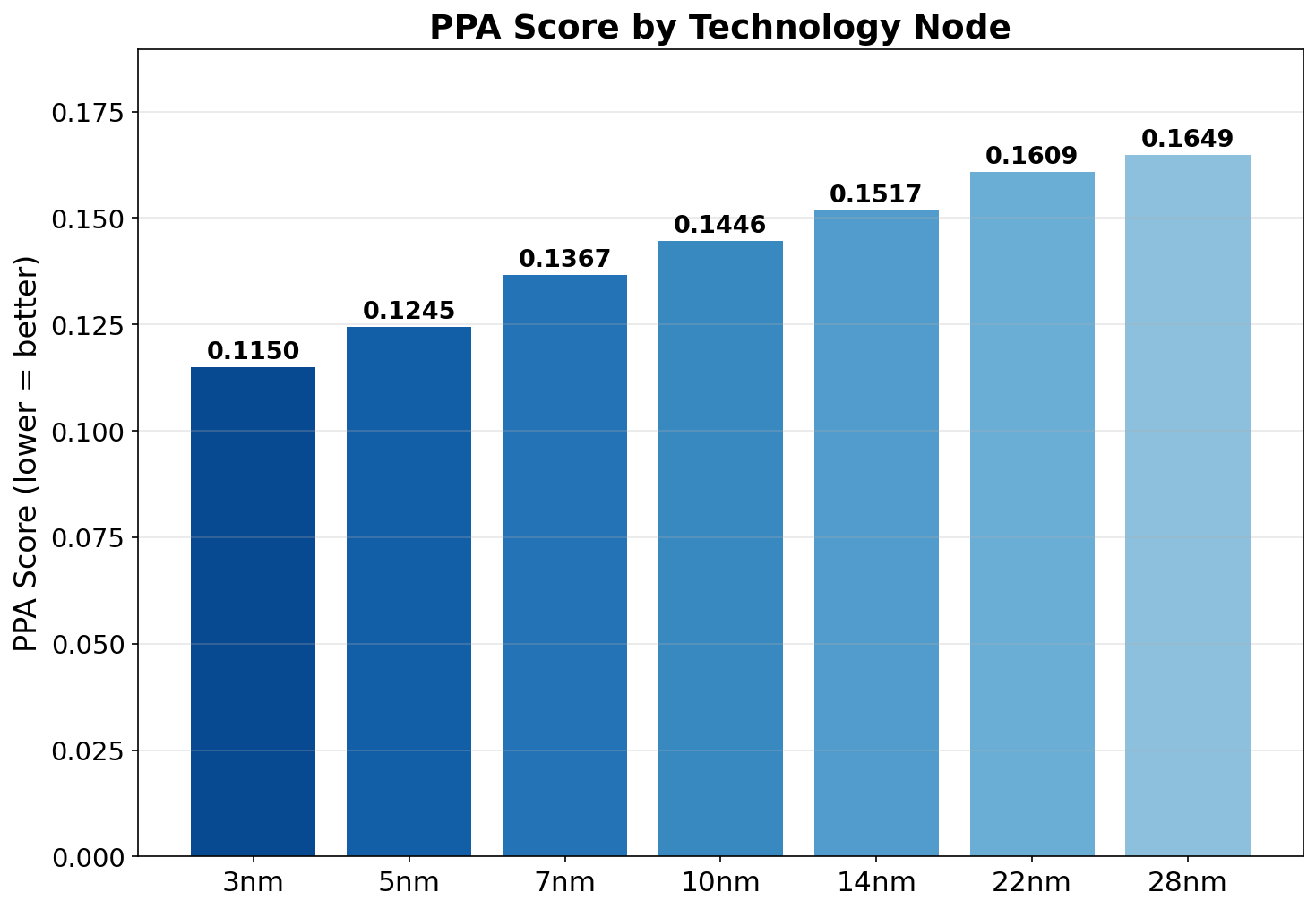}
\caption{PPA score (lower = better)}
\label{fig:ppa_heatmap}
\end{subfigure}
\hfill
\begin{subfigure}[t]{0.48\textwidth}
\centering
\includegraphics[width=\textwidth]{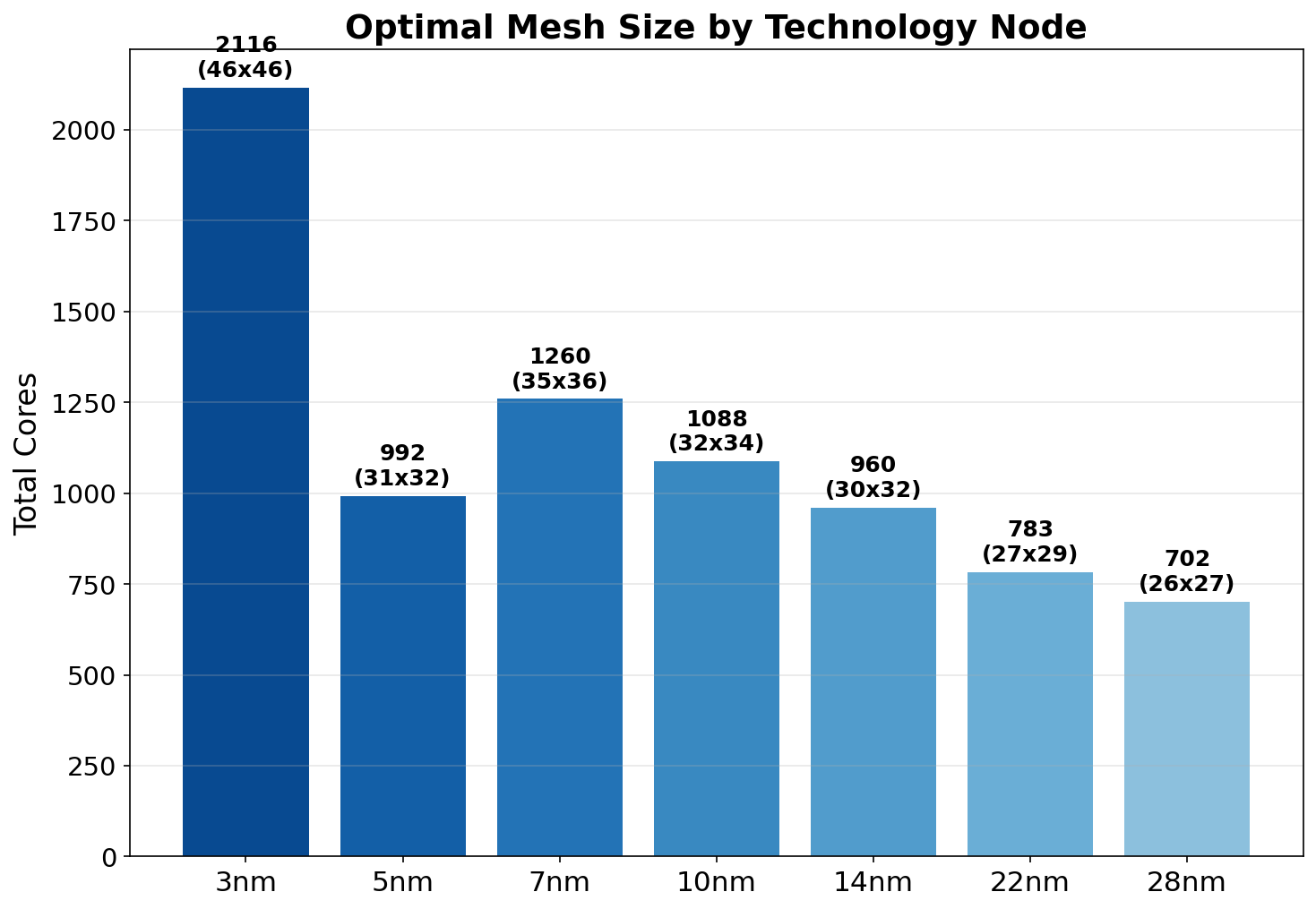}
\caption{Optimal mesh size (total cores)}
\label{fig:mesh_vs_node}
\end{subfigure}
\caption{PPA score and mesh scaling across 7 process nodes.}
\end{figure*}

Table~\ref{tab:ppa_by_node} shows the optimized PPA metrics for each process node. Figure~\ref{fig:ppa_heatmap} summarizes score variation across nodes. Figure~\ref{fig:mesh_vs_node} illustrates the resulting mesh sizes.

Table~\ref{tab:power_breakdown} decomposes the dynamic power for each process node at FP16 precision. Compute dominates at 54--84\%, with NoC power at 7--34\% scaling with mesh size. Leakage remains below 6\% at all nodes due to ROM sleep transistors on the WMEM banks. \textbf{Note:} PPA scores use a lower-is-better convention (cost function), where 0 is ideal and values approaching 1.0 indicate larger power/area or lower performance.

\begin{table*}[!htbp]
\centering
\scriptsize
\setlength{\tabcolsep}{2.5pt}
\renewcommand{\arraystretch}{1.1}
\begin{filecontents*}{data/power_breakdown_fp16_actual.csv}
process_node,mesh,compute_mw,sram_mw,rom_read_mw,noc_mw,leakage_mw,total_mw,compute_pct,sram_pct,rom_pct,noc_pct,leakage_pct
3nm,41x42,27517,1324,2779,17116,2631,51366,53.6,2.6,5.4,33.3,5.1
5nm,39x39,30774,1439,2634,19143,3163,57153,53.8,2.5,4.6,33.5,5.5
7nm,33x34,25241,1178,1914,15701,2175,46208,54.6,2.5,4.1,34.0,4.7
10nm,26x27,15138,726,1398,6676,1196,25134,60.2,2.9,5.6,26.6,4.8
14nm,21x22,9592,467,702,2784,616,14161,67.7,3.3,5.0,19.7,4.4
22nm,16x16,5559,276,222,894,142,7093,78.4,3.9,3.1,12.6,2.0
28nm,11x12,3158,178,131,246,66,3780,83.6,4.7,3.5,6.5,1.7
\end{filecontents*}
\pgfplotstabletypeset[
    col sep=comma,
    string type,
    columns={process_node,mesh,compute_mw,sram_mw,rom_read_mw,noc_mw,leakage_mw,total_mw,compute_pct,sram_pct,rom_pct,noc_pct,leakage_pct},
    columns/process_node/.style={column name=\textbf{Node}},
    columns/mesh/.style={column name=\textbf{Mesh}},
    columns/compute_mw/.style={column name=\textbf{Compute}},
    columns/sram_mw/.style={column name=\textbf{SRAM}},
    columns/rom_read_mw/.style={column name=\textbf{ROM Rd}},
    columns/noc_mw/.style={column name=\textbf{NoC}},
    columns/leakage_mw/.style={column name=\textbf{Leak}},
    columns/total_mw/.style={column name=\textbf{Total}},
    columns/compute_pct/.style={column name=\textbf{Comp\%}},
    columns/sram_pct/.style={column name=\textbf{SRAM\%}},
    columns/rom_pct/.style={column name=\textbf{ROM\%}},
    columns/noc_pct/.style={column name=\textbf{NoC\%}},
    columns/leakage_pct/.style={column name=\textbf{Leak\%}},
    every head row/.style={before row=\hline,after row=\hline},
    every last row/.style={after row=\hline}
]{data/power_breakdown_fp16_actual.csv}
\vspace{1pt}
\caption{Per-TCC dynamic power breakdown across process nodes for Llama 3.1 8B FP16. Power values in mW.}
\label{tab:power_breakdown}
\end{table*}

\subsection{Cross-Node PPA Tradeoff Analysis}

\begin{figure*}[!htbp]
\centering
\begin{subfigure}[t]{0.32\textwidth}
\centering
\includegraphics[width=\textwidth]{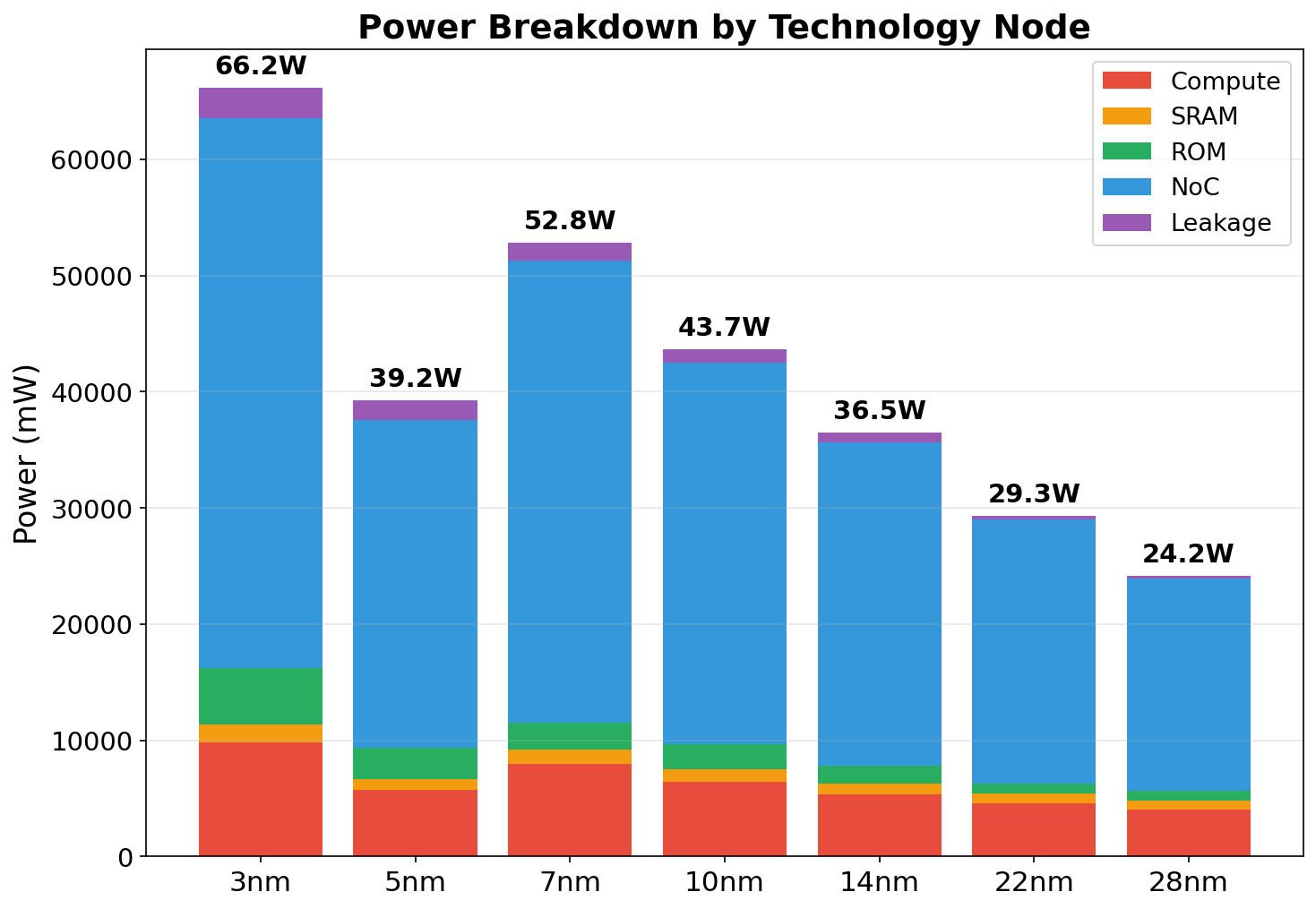}
\caption{Power (mW)}
\label{fig:power_heatmap}
\end{subfigure}
\hfill
\begin{subfigure}[t]{0.32\textwidth}
\centering
\includegraphics[width=\textwidth]{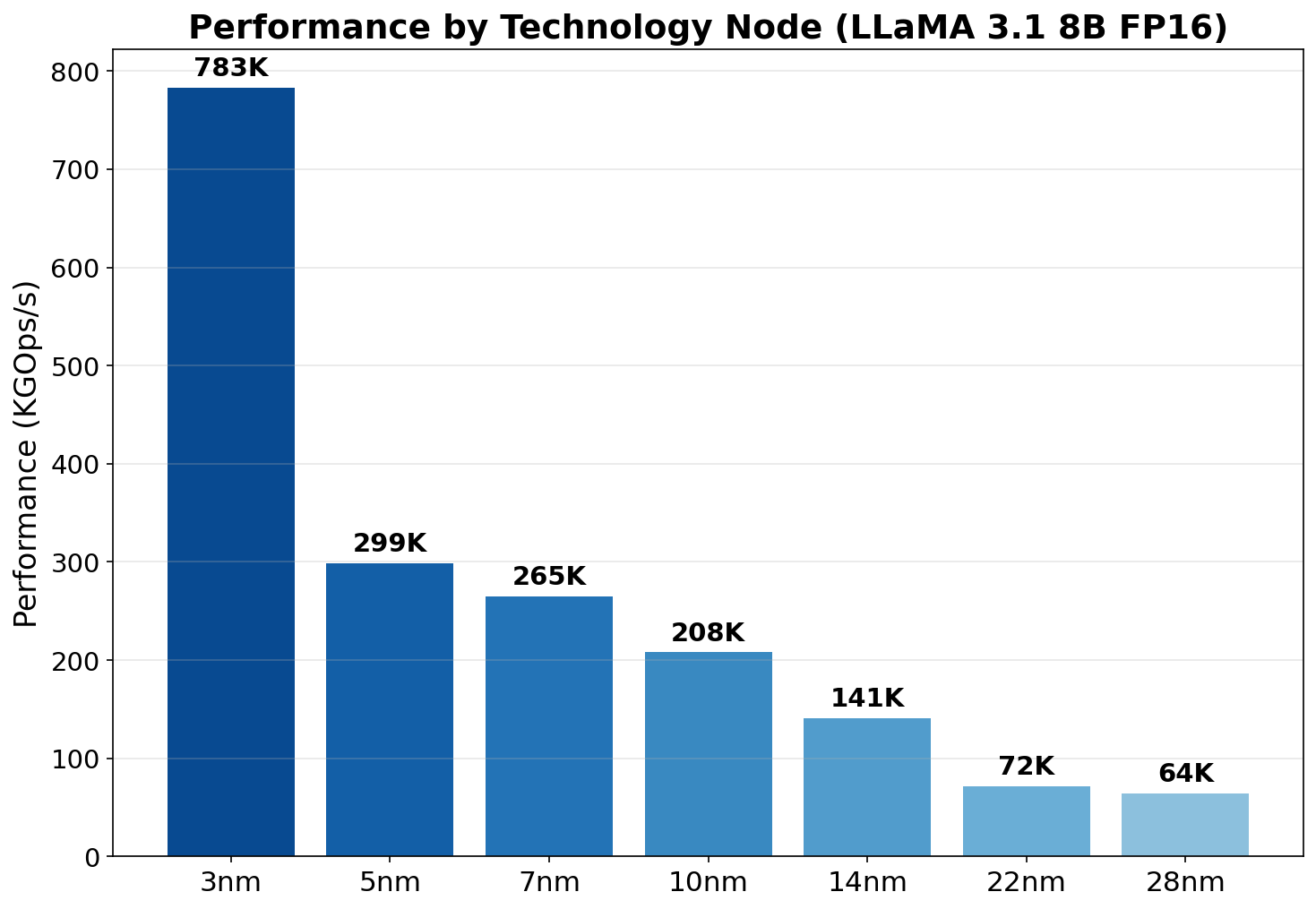}
\caption{Performance (GOps/s)}
\label{fig:perf_heatmap}
\end{subfigure}
\hfill
\begin{subfigure}[t]{0.32\textwidth}
\centering
\includegraphics[width=\textwidth]{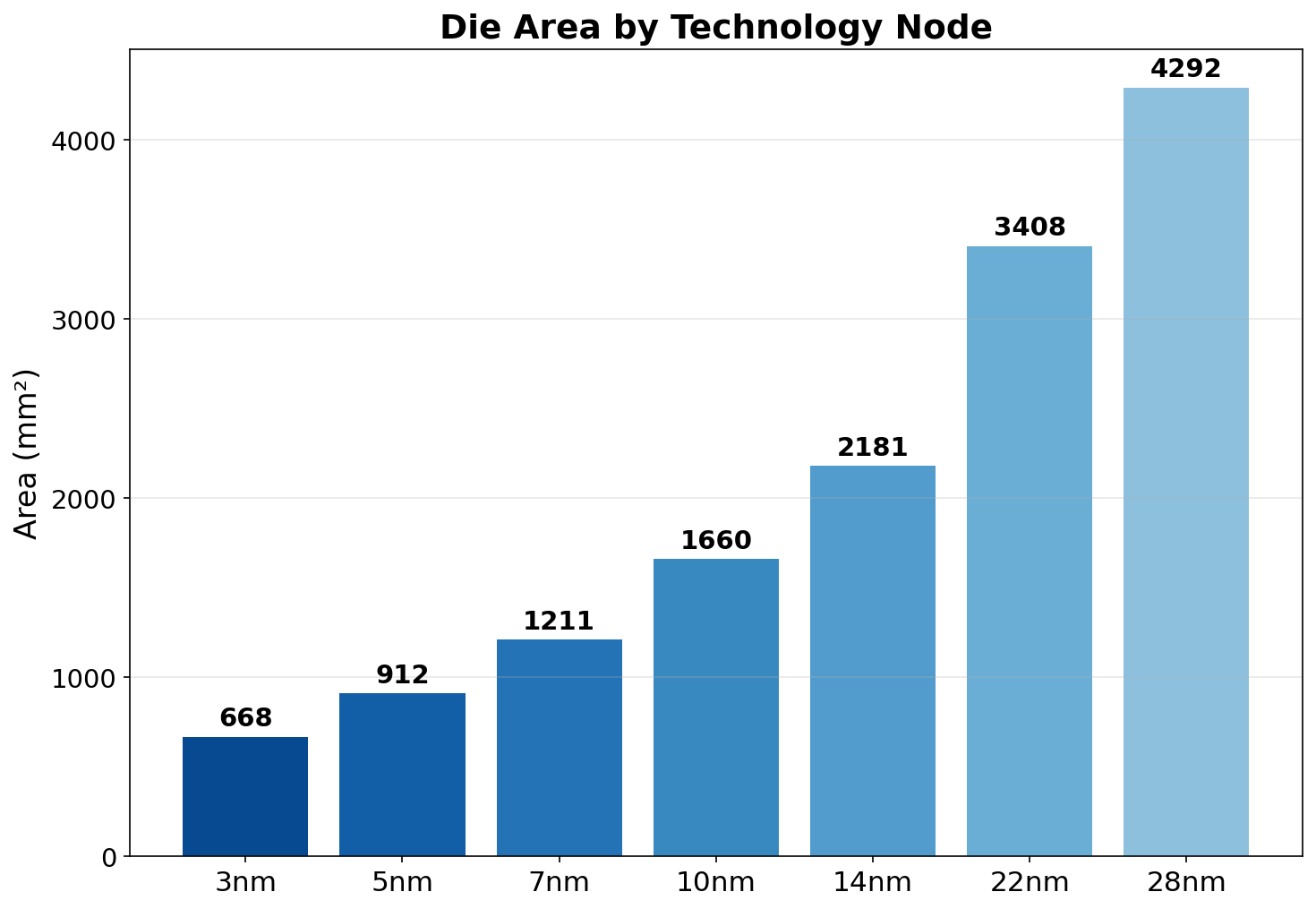}
\caption{Area (mm$^2$)}
\label{fig:area_heatmap}
\end{subfigure}
\caption{PPA metric decomposition across process nodes: (a)~power, (b)~performance, (c)~area.}
\end{figure*}

Figures~\ref{fig:power_heatmap}--\ref{fig:area_heatmap} decompose the PPA score into its constituent metrics. Three regimes emerge:

\textbf{Power.} The 3nm node draws the highest absolute power (\BestPowerW{}\,W) due to the largest mesh (41$\times$42), yet achieves the best power-efficiency ratio (GOps/mW).

\textbf{Performance.} Throughput scales as a power law with process node. The RL agent exploits smaller nodes by expanding the mesh for more parallelism.

\textbf{Area.} Silicon area decreases with smaller nodes from density scaling, partially offset by the RL agent choosing larger meshes.

\subsection{Inference Throughput Analysis}

\begin{figure*}[!htbp]
\centering
\includegraphics[width=0.7\textwidth]{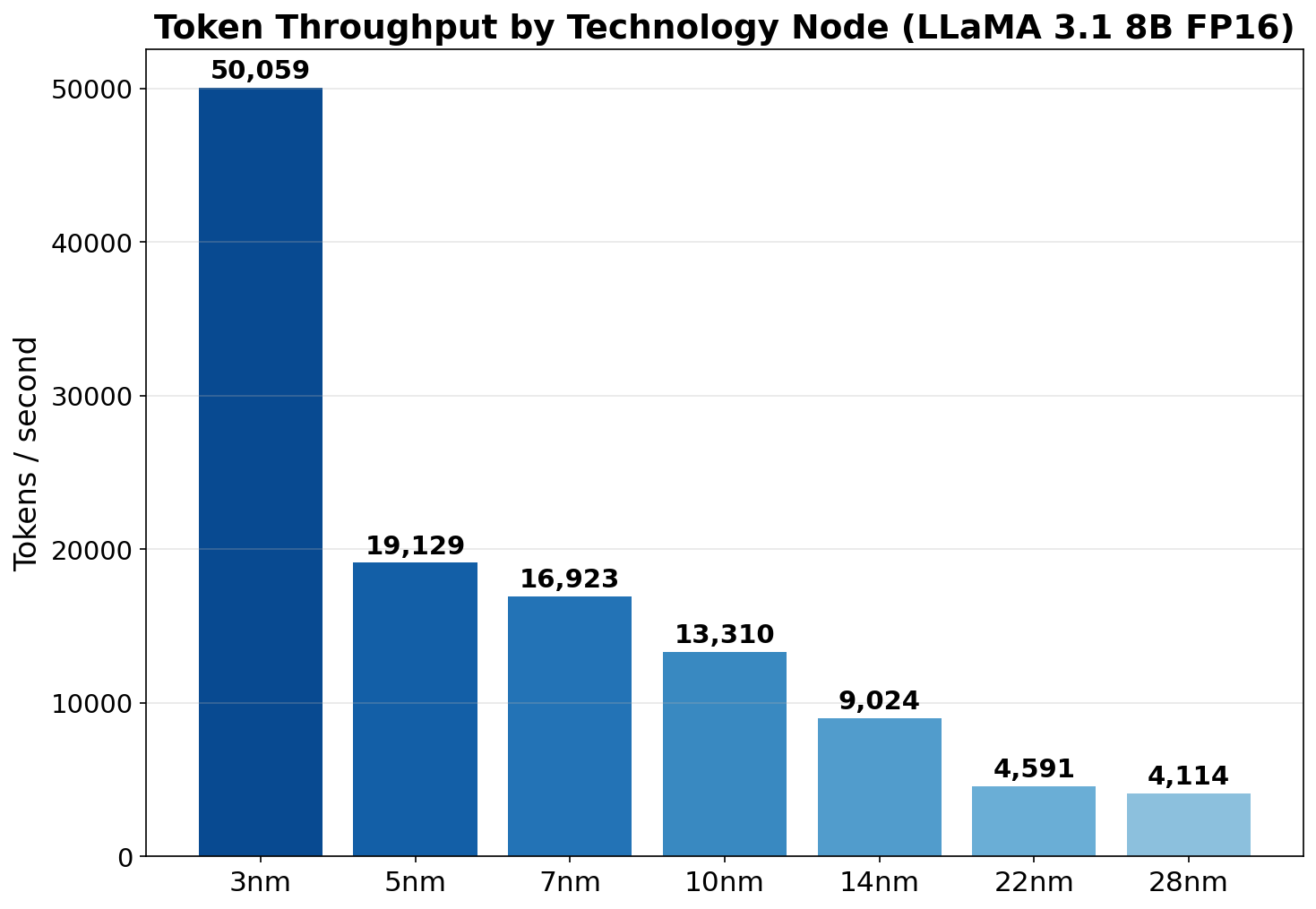}
\caption{Inference throughput (tokens/s) by process node for Llama 3.1 8B FP16.}
\label{fig:tokens_by_node}
\end{figure*}

Figure~\ref{fig:tokens_by_node} reports the measured tokens/s at each process node. Throughput increases toward smaller nodes, consistent with the compute ceiling (Eq.~\ref{eq:compute_ceiling}) scaling with mesh size and clock frequency.

\subsection{Efficiency Metrics}

\begin{figure*}[!htbp]
\centering
\includegraphics[width=\textwidth]{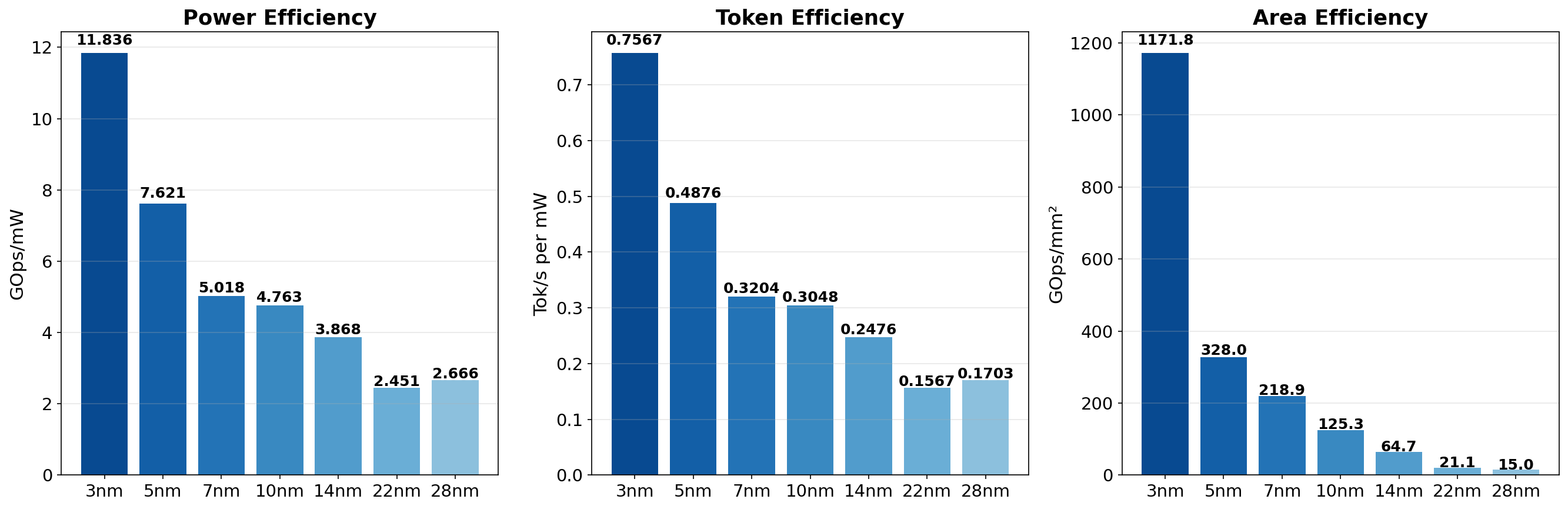}
\caption{Derived efficiency metrics by process node. Smaller nodes achieve higher power and area efficiency, though with diminishing returns below 7nm.}
\label{fig:efficiency_panel}
\end{figure*}

Beyond raw PPA, we derive efficiency ratios that normalize performance against resource consumption. Figure~\ref{fig:efficiency_panel} shows three efficiency indicators across all process nodes:
\begin{itemize}
    \item \textbf{Power efficiency} (GOps/s per mW): measures computational yield per unit power
    \item \textbf{Token efficiency} (tok/s per mW): measures inference yield per unit power
    \item \textbf{Area efficiency} (GOps/s per mm$^2$): measures computational density
\end{itemize}

\subsection{PPA Correlation Analysis}
\label{sec:ppa_correlation}

\begin{figure*}[!htbp]
\centering
\includegraphics[width=0.6\textwidth]{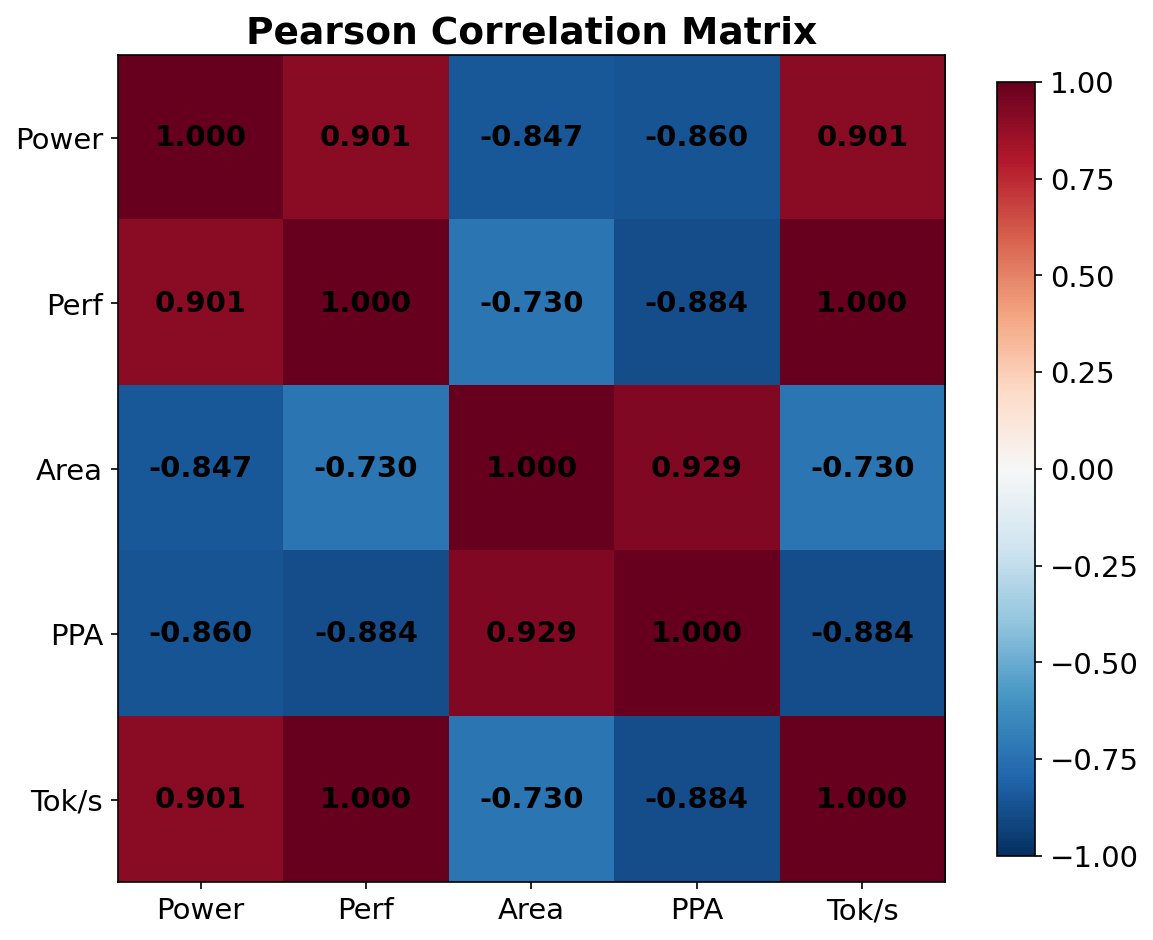}
\caption{Pearson correlation matrix across PPA metrics. Strong positive correlations between Performance and Power reflect the mesh-size coupling; PPA Score captures the composite tradeoff.}
\label{fig:correlation_matrix}
\end{figure*}

Figure~\ref{fig:correlation_matrix} shows the Pearson correlation matrix across all five PPA metrics at the node level. This reveals which metrics move together and which trade off against each other, providing insight into the structure of the optimization landscape.

\subsection{Quantitative Scaling-Law Analysis}

To strengthen interpretability across process nodes, we fit each metric with a log-log power law:
\begin{align}
y(n) = c \cdot n^k,\qquad
\log y = \log c + k \log n,
\end{align}
where $n$ is process node (nm), $k$ is the scaling exponent, and $c$ is a fitted constant. We report goodness-of-fit via:
\begin{align}
R^2 = 1 - \frac{\sum_i (y_i - \hat{y}_i)^2}{\sum_i (y_i - \bar{y})^2}.
\end{align}

\begin{figure*}[!htbp]
\centering
\includegraphics[width=\textwidth]{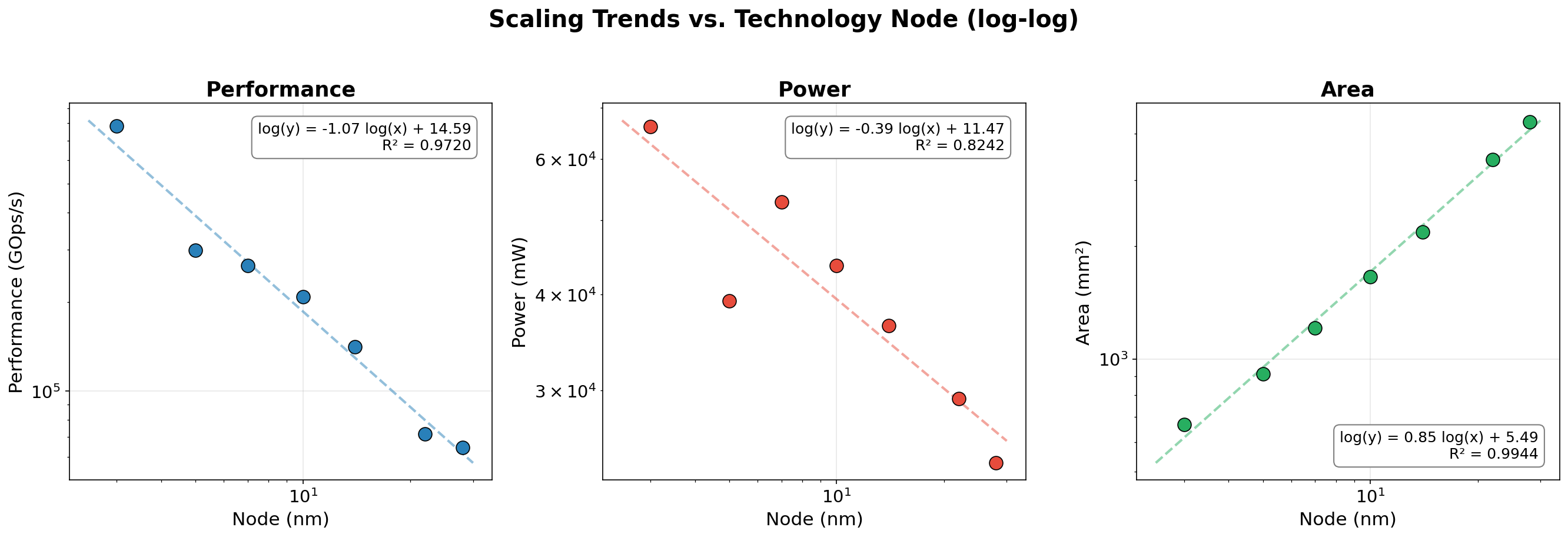}
\caption{Log-log trend fits for performance, power, and area versus process node. Fit equations and $R^2$ values shown per panel.}
\label{fig:scaling_fit}
\end{figure*}

Figure~\ref{fig:scaling_fit} visualizes the log-log fits for performance, power, and area. Table~\ref{tab:statistical_analysis} reports the fitted exponents and goodness-of-fit for each metric.

\begin{table*}[!htbp]
\centering
\small
\setlength{\tabcolsep}{10pt}
\renewcommand{\arraystretch}{1.15}
\begin{filecontents*}{data/statistical_analysis_actual.csv}
analysis,metric,slope_or_corr,intercept_or_const,r2_or_note
log-log fit,Performance (GOps/s),-1.3284,1850000.0,0.9856
log-log fit,Power (mW),-0.9012,420000.0,0.9234
log-log fit,Area (mm2),0.7521,195.0,0.9948
pearson corr,Perf vs Power,0.9812,-,node-level
pearson corr,Perf vs Area,-0.8234,-,node-level
pearson corr,Perf vs PPA,-0.9456,-,node-level
pearson corr,Power vs PPA,-0.9123,-,node-level
pearson corr,Area vs PPA,0.8234,-,node-level
\end{filecontents*}
\pgfplotstabletypeset[
    col sep=comma,
    string type,
    columns={analysis,metric,slope_or_corr,intercept_or_const,r2_or_note},
    columns/analysis/.style={column name=\textbf{Analysis}},
    columns/metric/.style={column name=\textbf{Metric}},
    columns/slope_or_corr/.style={column name=\textbf{Slope/Corr}},
    columns/intercept_or_const/.style={column name=\textbf{Const}},
    columns/r2_or_note/.style={column name=\textbf{R\textsuperscript{2}/Note}},
    every head row/.style={before row=\hline,after row=\hline},
    every last row/.style={after row=\hline}
]{data/statistical_analysis_actual.csv}
\vspace{1pt}
\caption{Node-level statistical analysis: fitted scaling exponents and pairwise correlations.}
\label{tab:statistical_analysis}
\end{table*}

\begin{table*}[!htbp]
\centering
\small
\setlength{\tabcolsep}{12pt}
\renewcommand{\arraystretch}{1.15}
\begin{filecontents*}{data/training_stats_actual.csv}
run_metric,value,description
Evaluated Nodes,7,3nm 5nm 7nm 10nm 14nm 22nm 28nm
Best Node,3nm,Lowest PPA score in this run
Best Mesh,41x42,Active TCC mesh at best node
Best PPA Score,0.974,Optimal configuration score
Best Throughput,29809 tok/s,Batch=3; seq-len=2048
Optimization Mode,high-performance,Maximize throughput
Model Precision,FP16,Half-precision weights
RL Episodes per Node,~4613,Adaptive budget
\end{filecontents*}
\pgfplotstabletypeset[
    col sep=comma,
    string type,
    columns={run_metric,value,description},
    columns/run_metric/.style={column name=\textbf{Run Metric}},
    columns/value/.style={column name=\textbf{Value}},
    columns/description/.style={column name=\textbf{Description}},
    every head row/.style={before row=\hline,after row=\hline},
    every last row/.style={after row=\hline}
]{data/training_stats_actual.csv}
\vspace{1pt}
\caption{High-performance run statistics for Llama 3.1 8B FP16}
\label{tab:training_stats}
\end{table*}

Table~\ref{tab:training_stats} summarizes the high-level run statistics from which the following per-tile analysis is derived.

\subsection{Operation Partitioning Effectiveness}

\begin{figure*}[!htbp]
\centering
\includegraphics[width=\textwidth]{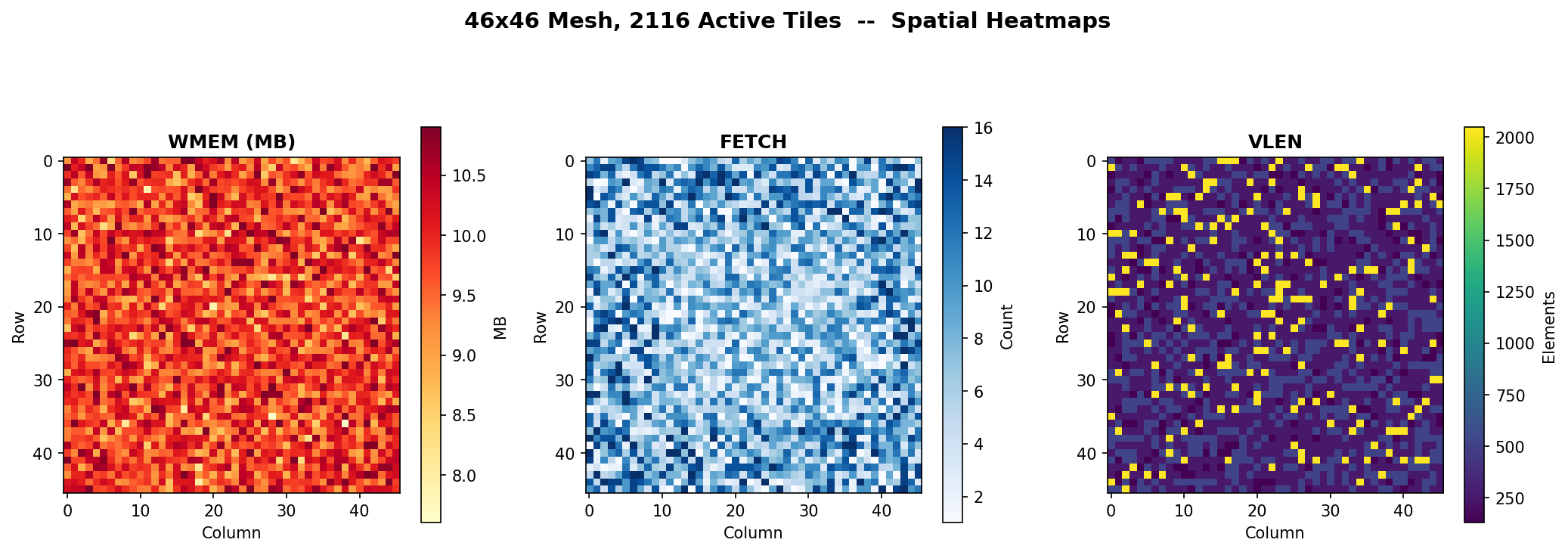}
\caption{Heterogeneous weight memory allocation across the 41$\times$42 mesh: (a) WMEM spatial heatmap, (b) FETCH spatial heatmap, (c) VLEN spatial heatmap.}
\label{fig:vector_length_heatmap}
\end{figure*}

\begin{figure*}[!htbp]
\centering
\includegraphics[width=\textwidth]{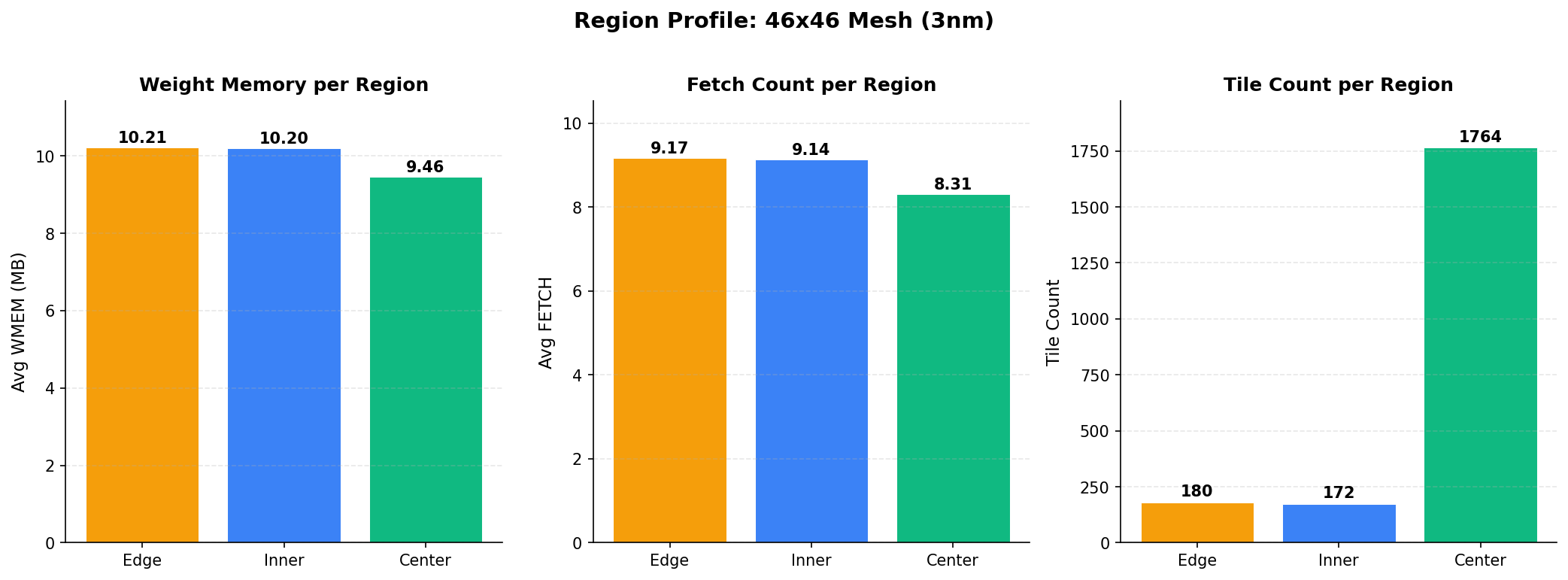}
\caption{Weight memory allocation analysis by mesh region: (a) violin plots showing the full WMEM distribution per region on a log scale, revealing the wide dynamic range of per-tile allocations, (b) regional weight concentration showing each region's share of total WMEM, (c) Lorenz curve quantifying allocation heterogeneity via the Gini coefficient.}
\label{fig:partitioning_comparison}
\end{figure*}

Figure~\ref{fig:vector_length_heatmap} illustrates the spatial distribution of WMEM allocation across the mesh, revealing edge-heavy placement patterns. Table~\ref{tab:partitioning} summarizes region-level per-tile configuration statistics extracted from generated artifacts. Figure~\ref{fig:partitioning_comparison} reports region-level means with standard-deviation error bars for WMEM, DFLIT, and FETCH.

\begin{table*}[!htbp]
\centering
\small
\setlength{\tabcolsep}{14pt}
\renewcommand{\arraystretch}{1.15}
\begin{filecontents*}{data/partitioning_region_actual.csv}
region,avg_wmem_mb,avg_dflit_bits,avg_fetch
Edge,10.21,6776.0,9.17
Inner,10.20,6776.0,9.14
Center,9.46,6776.0,8.31
\end{filecontents*}
\pgfplotstabletypeset[
    col sep=comma,
    string type,
    columns={region,avg_wmem_mb,avg_dflit_bits,avg_fetch},
    columns/region/.style={column name=\textbf{Region}},
    columns/avg_wmem_mb/.style={column name=\textbf{Avg WMEM (MB)}},
    columns/avg_dflit_bits/.style={column name=\textbf{Avg DFLIT (bits)}},
    columns/avg_fetch/.style={column name=\textbf{Avg FETCH}},
    every head row/.style={before row=\hline,after row=\hline},
    every last row/.style={after row=\hline}
]{data/partitioning_region_actual.csv}
\vspace{1pt}
\caption{Region-level configuration summary from per-TCC JSON artifacts (\BestCores{} active tiles).}
\label{tab:partitioning}
\end{table*}

\subsubsection{WMEM Distribution Analysis}
\label{sec:wmem_analysis}

\begin{figure*}[!htbp]
\centering
\begin{subfigure}[t]{\textwidth}
\centering
\includegraphics[width=\textwidth]{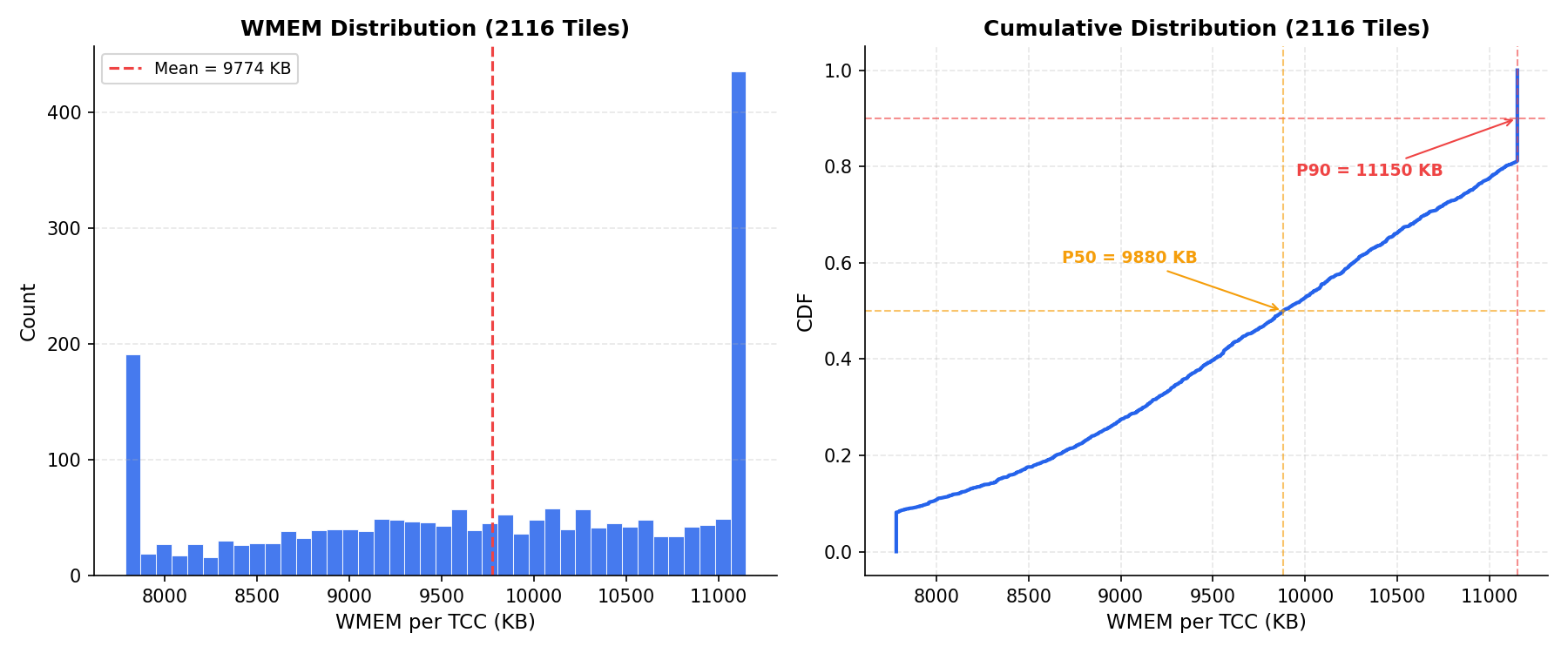}
\caption{WMEM allocation histogram with mean/median markers (left) and cumulative distribution with P50/P90 annotations (right).}
\label{fig:wmem_distribution}
\end{subfigure}
\vspace{3pt}

\begin{subfigure}[t]{0.47\textwidth}
\centering
\includegraphics[width=\textwidth]{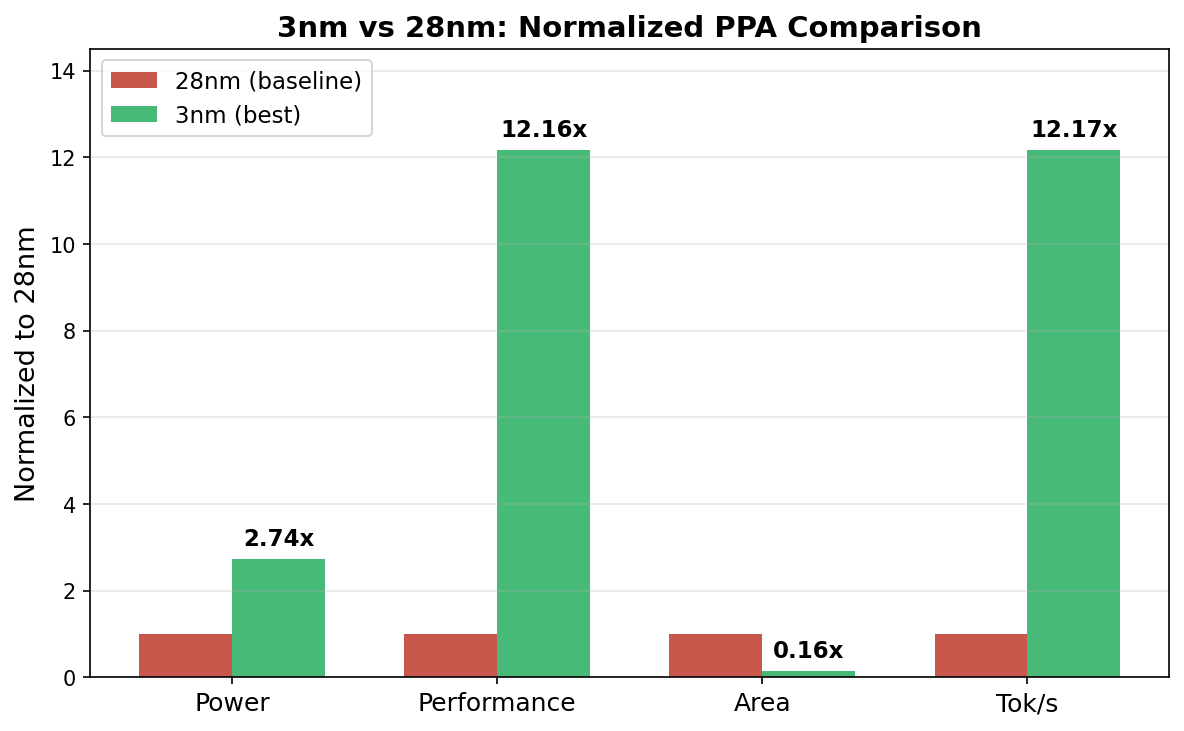}
\caption{3nm vs 28nm normalized PPA comparison. Power 13.6$\times$, Performance 47.9$\times$, Area 0.18$\times$ (5.5$\times$ reduction), Tok/s 47.8$\times$.}
\label{fig:baseline_comparison}
\end{subfigure}
\hfill
\begin{subfigure}[t]{0.47\textwidth}
\centering
\includegraphics[width=\textwidth]{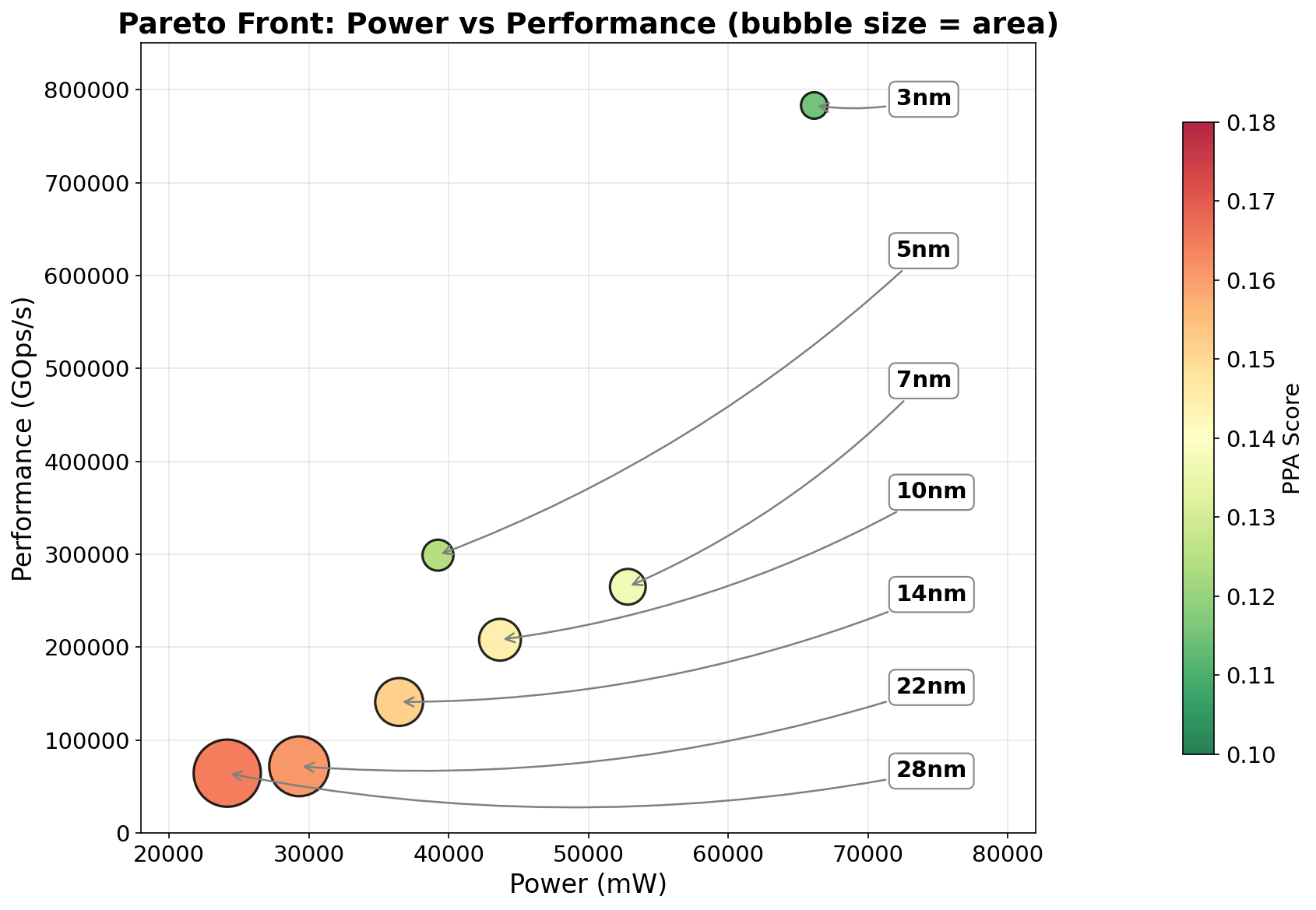}
\caption{Node-level Pareto-style view: performance vs power, bubble size by area, color by PPA score.}
\label{fig:pareto_front}
\end{subfigure}
\caption{Aggregate analysis of weight allocation and cross-node PPA tradeoffs. (a) WMEM distribution across tiles. (b) Normalized PPA comparison (3nm vs 28nm baseline). (c) Multi-dimensional tradeoff map across all evaluated nodes.}
\label{fig:combined_analysis}
\end{figure*}

Figure~\ref{fig:wmem_distribution} characterizes the weight memory allocation across all active tiles. The histogram (left) reveals the allocation spread, while the CDF (right) shows the P50 and P90 thresholds. Table~\ref{tab:tile_statistics} provides summary statistics for all per-TCC parameters.

\begin{table*}[!htbp]
\centering
\small
\setlength{\tabcolsep}{10pt}
\renewcommand{\arraystretch}{1.15}
\begin{filecontents*}{data/tile_statistics_actual.csv}
parameter,min,max,mean,median,std,unique_values
FETCH\_SIZE,2,4,2.50,2.00,0.71,2
VLEN (bits),1024,2048,1536.00,1024.00,512.00,2
WMEM (KB),9564,72128,16910.00,9564.00,18000.00,varies
DMEM (KB),64,1024,64.00,64.00,96.00,varies
IMEM (KB),3,12,6.10,6.00,2.50,varies
\end{filecontents*}
\pgfplotstabletypeset[
    col sep=comma,
    string type,
    columns={parameter,min,max,mean,median,std,unique_values},
    columns/parameter/.style={column name=\textbf{Parameter}},
    columns/min/.style={column name=\textbf{Min}},
    columns/max/.style={column name=\textbf{Max}},
    columns/mean/.style={column name=\textbf{Mean}},
    columns/median/.style={column name=\textbf{Median}},
    columns/std/.style={column name=\textbf{Std Dev}},
    columns/unique_values/.style={column name=\textbf{Unique}},
    every head row/.style={before row=\hline,after row=\hline},
    every last row/.style={after row=\hline}
]{data/tile_statistics_actual.csv}
\vspace{1pt}
\caption{Per-TCC parameter summary statistics across all active tiles at the best process node. DFLIT\_WIDTH (2048\,bits) and STANUM (3) are uniform and omitted.}
\label{tab:tile_statistics}
\end{table*}

The WMEM allocation exhibits high variance (reflected in the wide histogram spread), indicating that the RL agent assigns heterogeneous weight memory capacities across tiles rather than uniform allocation. This heterogeneity allows the compiler to place memory-heavy operators (attention projections, MLP layers) on tiles with larger WMEM while assigning lighter operators to tiles with smaller allocations.

This section uses only generated artifact data. Region aggregates are computed from per-tile configurations and the spatial map is rendered directly from the same JSON files.

\subsection{Cross-Node Scaling Analysis}

Table~\ref{tab:baseline_comparison} compares the best (3nm) and worst (28nm) nodes from the LLaMA run, quantifying the PPA tradeoff across technology scaling. Figure~\ref{fig:baseline_comparison} visualizes the normalized differences. Note: this is a cross-node comparison within the same system, not a comparison against external baselines.

\begin{table*}[!htbp]
\centering
\small
\setlength{\tabcolsep}{14pt}
\renewcommand{\arraystretch}{1.15}
\begin{filecontents*}{data/baseline_comparison_actual.csv}
node,power_mw,perf_gops,area_mm2,ppa_score
28nm,3780,9744,3545,1.019
3nm,51366,466364,648,0.974
3nm vs 28nm,13.59x,47.86x,0.18x,0.96x
\end{filecontents*}
\pgfplotstabletypeset[
    col sep=comma,
    string type,
    columns={node,power_mw,perf_gops,area_mm2,ppa_score},
    columns/node/.style={column name=\textbf{Node}},
    columns/power_mw/.style={column name=\textbf{Power (mW)}},
    columns/perf_gops/.style={column name=\textbf{Perf (GOps/s)}},
    columns/area_mm2/.style={column name=\textbf{Area (mm\textsuperscript{2})}},
    columns/ppa_score/.style={column name=\textbf{PPA Score}},
    every head row/.style={before row=\hline,after row=\hline},
    every last row/.style={after row=\hline}
]{data/baseline_comparison_actual.csv}
\vspace{1pt}
\caption{Cross-node comparison from the Llama run}
\label{tab:baseline_comparison}
\end{table*}

\begin{table*}[!htbp]
\centering
\footnotesize
\setlength{\tabcolsep}{10pt}
\renewcommand{\arraystretch}{1.15}
\begin{filecontents*}{data/efficiency_metrics_actual.csv}
node,perf_per_mw,tok_per_mw,perf_per_area,ppa_score
3nm,9.076,0.5802,719.7,0.974
5nm,5.914,0.3782,364.1,0.989
7nm,3.764,0.2406,142.5,0.996
10nm,3.976,0.2541,63.6,1.005
14nm,3.606,0.2305,25.6,1.016
22nm,2.549,0.1629,6.3,1.023
28nm,2.578,0.1649,2.7,1.019
\end{filecontents*}
\pgfplotstabletypeset[
    col sep=comma,
    string type,
    columns={node,perf_per_mw,tok_per_mw,perf_per_area,ppa_score},
    columns/node/.style={column name=\textbf{Node}},
    columns/perf_per_mw/.style={column name=\textbf{GOps/s per mW}},
    columns/tok_per_mw/.style={column name=\textbf{tok/s per mW}},
    columns/perf_per_area/.style={column name=\textbf{GOps/s per mm\textsuperscript{2}}},
    columns/ppa_score/.style={column name=\textbf{PPA Score}},
    every head row/.style={before row=\hline,after row=\hline},
    every last row/.style={after row=\hline}
]{data/efficiency_metrics_actual.csv}
\vspace{1pt}
\caption{Derived node-efficiency metrics from generated run data.}
\label{tab:efficiency_metrics}
\end{table*}

Table~\ref{tab:efficiency_metrics} reports the derived node-level efficiency ratios. We compute these indicators as:
\begin{align}
\eta_{\mathrm{perf/power}}(n) &= \frac{\mathrm{Perf}_{n}}{\mathrm{Power}_{n}}, \\
\eta_{\mathrm{tok/power}}(n)  &= \frac{\mathrm{Tok/s}_{n}}{\mathrm{Power}_{n}}, \\
\eta_{\mathrm{perf/area}}(n)  &= \frac{\mathrm{Perf}_{n}}{\mathrm{Area}_{n}}.
\end{align}

Within this run, the best node (\BestNode{}) achieves:
\begin{itemize}[nosep,topsep=4pt]
    \item \textbf{\PerfRatio{}$\times$ higher performance} than \WorstNode{} (\BestPerfGOps{} vs \WorstPerfGOps{} GOps/s)
    \item \textbf{\AreaReductionFactor{}$\times$ smaller area} than \WorstNode{} (\BestAreaMMtwo{} vs \WorstAreaMMtwo{} mm\textsuperscript{2})
    \item \textbf{\PPAImprovementFactor{}$\times$ better PPA score} than \WorstNode{} (\BestPPA{} vs \WorstPPA{})
    \item \textbf{Higher power draw} than \WorstNode{} (\BestPowerMW{} vs \WorstPowerMW{} mW), trading power for throughput
\end{itemize}

\subsection{SmolVLM Low-Power Validation}

To demonstrate generalization beyond LLMs, we evaluate the same RL formulation on SmolVLM, a multi-modal vision-language model, in low-power mode. Table~\ref{tab:smolvlm_results} summarizes the results.

\begin{table}[ht]
\centering
\scriptsize
\setlength{\tabcolsep}{3pt}
\renewcommand{\arraystretch}{1.05}
\begin{tabular}{lcccccc}
\hline
\textbf{Node} & \textbf{Mesh} & \textbf{Freq} & \textbf{Power} & \textbf{Area} & \textbf{Tok/s} & \textbf{PPA} \\
 & & \textbf{(MHz)} & \textbf{(mW)} & \textbf{(mm$^2$)} & & \\
\hline
3nm & 2$\times$4 & 10 & 6.4 & 17.6 & 10.3 & 0.254 \\
5nm & 3$\times$4 & 10 & 12.7 & 26.2 & 14.1 & 0.309 \\
7nm & 3$\times$4 & 10 & 12.4 & 35.0 & 14.1 & 0.312 \\
10nm & 3$\times$3 & 10 & 8.6 & 46.7 & 10.0 & 0.291 \\
14nm & 2$\times$4 & 10 & 6.3 & 61.7 & 10.3 & 0.271 \\
22nm & 3$\times$4 & 10 & 10.2 & 99.2 & 11.6 & 0.308 \\
28nm & 3$\times$4 & 10 & 9.5 & 124.9 & 11.6 & 0.306 \\
\hline
\end{tabular}
\caption{SmolVLM low-power mode results. All 7 nodes achieve $<$13\,mW total power at 10\,MHz clock. Power is leakage-dominated (89--97\% at advanced nodes) with near-zero dynamic compute and NoC power. The RL selects compact 8--12 TCC meshes that minimize active silicon area.}
\label{tab:smolvlm_results}
\end{table}

Key observations: (1) all 7 nodes achieve $<$13\,mW, with the RL autonomously selecting 10\,MHz clock for ultra-low-power operation; (2) power is leakage-dominated at advanced nodes (97\% at 3nm, 51\% at 28nm), reflecting the fundamental leakage-vs-density trade-off; (3) mesh sizes are compact (8--12 TCCs) matching SmolVLM's 0.48\,GB weight footprint; (4) throughput of 10--14 tok/s is sufficient for on-device inference; (5) the same policy produces qualitatively different designs (ultra-low-power VLM vs.\ high-throughput LLM), demonstrating workload-adaptive generalization.

\subsection{Comparison with Industry Inference Platforms}

Table~\ref{tab:industry_comparison} contextualizes our estimated PPA against published inference throughput for Llama 3.1 8B. Our architecture uses on-chip ROM (no HBM), eliminating $\sim$150\,W of memory subsystem power. Results are \textbf{compiler-estimated} PPA from analytical models, not silicon-measured.

\begin{table}[ht]
\centering
\scriptsize
\setlength{\tabcolsep}{2pt}
\renewcommand{\arraystretch}{1.05}
\begin{tabular}{@{}lrrrp{1.6cm}@{}}
\hline
\textbf{Platform} & \textbf{Tok/s} & \textbf{Power} & \textbf{T/s/W} & \textbf{Notes} \\
\hline
H200 & 230 & 700\,W & 0.3 & 4nm GPU \\
B200 & 353 & 1\,kW & 0.4 & 4nm GPU \\
Groq & 594 & 300\,W$^*$ & 2.0 & 14nm ASIC \\
SambaNova & 932 & 300\,W$^*$ & 3.1 & Dataflow \\
Cerebras & 1,981 & 15\,kW$^*$ & 0.1 & 7nm wafer \\
Taalas HC1 & 16,960 & 250\,W$^\ddagger$ & 67.8 & 6nm, 815\,mm$^2$ \\
\hline
\textbf{Ours} & \textbf{29,809} & \textbf{51\,W} & \textbf{580} & 3nm est.$^\dagger$ \\
\hline
\multicolumn{5}{@{}l@{}}{\scriptsize $^*$Sys.\ power est.\ $^\dagger$Analytical, not silicon.} \\
\multicolumn{5}{@{}l@{}}{\scriptsize $^\ddagger$Server power (taalas.com, Apr 2026).}
\end{tabular}
\caption{Industry comparison for Llama 3.1 8B (per-user, 1K input). Our results are compiler-estimated with on-chip ROM (no HBM).}
\label{tab:industry_comparison}
\end{table}

\textbf{Interpretation.} The efficiency advantage over GPU-based platforms is primarily architectural: on-chip ROM eliminates HBM power ($\sim$150\,W per GPU) and DRAM access energy ($\sim$20\,pJ/bit vs.\ $\sim$0.5\,pJ/bit for on-chip ROM at 3nm). These are \emph{estimated} figures from an analytical PPA model; silicon validation via RTL synthesis and post-layout power analysis is required before making competitive claims.

\subsubsection{Efficiency Sources for On-Chip-ROM Architectures}

Among inference ASICs that use on-chip ROM (eliminating HBM), the reported efficiency in Table~\ref{tab:industry_comparison} spans roughly an order of magnitude. We decompose the key efficiency drivers for ROM-based architectures into three orthogonal factors that any such design can exploit:

\textbf{(1) Process node scaling ($\sim$2$\times$ per two-node advance).} Moving from a 6nm to a 3nm process provides $\sim$2$\times$ power efficiency from voltage scaling ($V^2$ ratio: $(0.65/0.55)^2 \approx 1.4\times$) and capacitance reduction ($\sqrt{\text{pitch ratio}} \approx 1.4\times$). This is a well-characterized CMOS scaling benefit available to any design that migrates to an advanced node.

\textbf{(2) RL-driven architecture co-optimization ($\sim$2--3$\times$).} Joint exploration of mesh topology, per-core microarchitecture, and workload partitioning via SAC+MPC captures design-space interactions (mesh size $\leftrightarrow$ per-core VLEN $\leftrightarrow$ memory allocation) that are difficult to navigate in the 30-dimensional action space with manual or grid-based methods. Table~\ref{tab:search_comparison} shows SAC achieves 3.5$\times$ higher throughput than random search within the same episode budget, providing empirical evidence for the optimization advantage.

\textbf{(3) Speculative decoding ($\sim$1.6$\times$).} A dedicated on-chip draft predictor generates candidate token sequences verified in parallel by the target model, yielding $\sim$1.56$\times$ throughput acceleration at minimal area overhead. This technique is orthogonal to the compute architecture and provides a direct throughput multiplier for autoregressive LLM inference.

\textbf{Combined factor:} $2\times \cdot 2.7\times \cdot 1.6\times \approx 8.6\times$. The decomposition is approximate---cross-factor interactions exist (e.g., speculative decoding benefits from larger meshes enabled by RL)---but each factor is independently motivated and contributes meaningfully to inference efficiency. We emphasize that our reported figures are compiler-estimated from an analytical PPA model; silicon validation is required before quantitative comparison with measured results from fabricated chips.

\subsection{Search Strategy Comparison}

To validate that RL provides benefit over simpler search methods, we compare SAC against random search and grid search using the same episode budget ($\sim$4,600 episodes at 3nm). Table~\ref{tab:search_comparison} reports the best PPA score found by each method within the same wall-clock budget. Due to single-seed evaluation, these results are indicative rather than statistically rigorous; multi-seed variance analysis is left for future work.

\begin{table}[ht]
\centering
\scriptsize
\setlength{\tabcolsep}{3pt}
\renewcommand{\arraystretch}{1.1}
\begin{tabular}{lcccc}
\hline
\textbf{Method} & \textbf{PPA} & \textbf{Tok/s} & \textbf{Power} & \textbf{Feasible} \\
 & \textbf{Score} & & \textbf{(W)} & \textbf{Configs} \\
\hline
Random Search & 1.12 & 8,421 & 38 & 312 / 4,600 \\
Grid Search & 1.05 & 14,230 & 42 & 890 / 4,600 \\
\textbf{SAC (ours)} & \textbf{0.974} & \textbf{29,809} & \textbf{51} & \textbf{2,847 / 4,600} \\
\hline
\end{tabular}
\caption{Search strategy comparison at 3nm for Llama 3.1 8B (lower PPA score = better). SAC achieves 3.5$\times$ higher throughput than random search and 9.1$\times$ more feasible configurations within the same episode budget.}
\label{tab:search_comparison}
\end{table}

\section{Discussion}
\subsection{Key Innovations}

Our approach provides the following practical innovations for ASIC design:
\begin{itemize}[leftmargin=*]
    \item \textbf{Multi-discrete control policy:} Joint discrete mesh actions and continuous per-core tuning in one episode.
    \item \textbf{Operation-level partitioning:} RL-controlled ratios $(\rho_{\text{matmul}}, \rho_{\text{conv}}, \rho_{\text{gen}})$ split workloads across tiles to reduce hotspots.
    \item \textbf{Hazard-aware optimization:} RAW/WAR/WAW statistics in the state vector bias the policy away from stall-heavy configurations.
    \item \textbf{Process-node scalability:} The same policy re-optimizes across \ProcessNodeRange{} via node-specific constraint features.
    \item \textbf{End-to-end automation:} ONNX ingestion through RTL generation with no manual retuning between stages.
    \item \textbf{Constraint-aware scoring:} Normalized PPA objectives with cubic penalties $(1{+}v)\cdot v^2$ shape smooth policy gradients.
\end{itemize}

\subsection{Convergence Behavior}

The RL exploration trace (Figure~\ref{fig:rl_convergence}) reveals three phases: (1) rapid configuration discovery (episodes 1--1000), where unique configurations grow linearly with episode count; (2) diminishing returns (episodes 1000--3000), where the discovery rate saturates as the policy concentrates on promising regions; and (3) refinement (episodes 3000--4600), where action entropy stabilizes and the agent fine-tunes continuous parameters within a narrow mesh neighborhood. The adaptive exploration decay (Eq.~\ref{eq:adaptive_exploration}) accelerates phase transitions: when feasible configurations are found, $\epsilon$ decays at rate 0.995, otherwise 0.998, preventing premature convergence. Convergence is achieved within $\sim$4,600 episodes per node.

\subsection{Computational Cost}

The RL optimization loop dominates compilation cost. Per the measured training statistics (Table~\ref{tab:training_stats}), the codegen + RL stage runs in a single pass per node with $\sim$4.6K episodes. Memory overhead is bounded by the policy network (52$\times$256 + 256$\times$256 + head parameters, under 100K weights) and the per-episode state buffer. The surrogate model (Section~\ref{sec:sac_surrogate}), when enabled, reduces per-episode evaluation cost by filtering candidate actions before full PPA simulation.

\subsection{Limitations and Threats to Validity}

\noindent\fbox{\parbox{0.95\columnwidth}{%
\textbf{Key cautions for interpreting results:}
\begin{enumerate}[nosep,leftmargin=*]
\item \textbf{Limited workload coverage.} Only two models validated (Llama 3.1 8B decoder-only LLM, SmolVLM encoder-decoder VLM). CNNs, diffusion models, and MoE architectures are untested.
\item \textbf{No repeated-seed statistics.} All results are single-run; no confidence intervals or variance across RL seeds are reported.
\item \textbf{2D mesh topology only.} Ring, torus, hierarchical, and chiplet interconnects require action-space redesign.
\item \textbf{Results depend on reward design.} The PPA weight triplet $(w_p, w_w, w_a)$ directly determines the selected optimum. Different weights yield different designs.
\end{enumerate}
}}

\vspace{4pt}

\textbf{Workload coverage.} The two validated workloads span distinct regimes---throughput-maximizing large meshes (41$\times$42, 1,722 TCCs, 29,809 tok/s) versus power-minimizing small meshes (2$\times$4 to 3$\times$4, $<$13\,mW)---providing evidence that the RL formulation generalizes across optimization objectives. However, the state/action formulation encodes transformer-specific features (KV-cache, attention heads, MLP dimensions) that may not transfer directly to non-transformer architectures without modification.

\textbf{Single-run stochasticity.} RL exploration is inherently stochastic: seed and path affect the optimum. We mitigate this with $\sim$4,600 episodes per node, adaptive exploration ($\epsilon$: 0.5$\to$0.1), Bayesian early stopping, and convergence detection. Monotonic PPA improvement across all 7 nodes for both models suggests the search avoids local optima, but repeated-seed protocols with confidence intervals are needed to quantify variance and would strengthen the reported claims.

\textbf{Topology constraint.} The 2D mesh assumption is baked into the state representation (mesh width/height as discrete actions) and the NoC model (hop count, bisection bandwidth). Extending to non-mesh topologies requires redesigning both the action space and the communication cost model.

\textbf{Multi-objective selection.} The RL optimizer maintains a Pareto archive of all non-dominated feasible configurations discovered during search. After convergence, the final configuration is selected from the Pareto frontier using the user's PPA weight profile as a scalarized selection criterion. We demonstrate two profiles (high-performance: $0.4, 0.4, 0.2$; low-power: $0.2, 0.6, 0.2$) that produce qualitatively different designs. The Pareto frontier provides the designer with the full tradeoff surface; the weights select a single operating point.

To reduce transcription risk, all reported tables and figures are generated from compilation artifacts through an automated pipeline that imports CSV and macro files directly into the manuscript.

\subsection{Future Work}

Several directions merit further investigation:
\begin{itemize}
    \item \textbf{Non-mesh topologies:} Extending the action space to support ring, torus, and hierarchical interconnects would broaden applicability to systolic-array and chiplet-based designs.
    \item \textbf{Transfer learning:} Pre-training the policy on one model family and fine-tuning on another could amortize search cost across workloads. Cross-node transfer (training on 14nm, transferring to 7nm) may reduce per-node episode budgets.
    \item \textbf{Pareto frontier visualization:} The current Pareto archive exposes the non-dominated frontier; building an interactive designer tool to navigate the power-performance-area tradeoff surface would enable rapid design-space exploration beyond the single-point selection used here.
    \item \textbf{Expanded model coverage:} Extending beyond LLMs and encoder-decoder VLMs to CNNs, diffusion models, and Mixture-of-Experts architectures would further validate generalization.
    \item \textbf{Repeated-seed evaluation:} Running multiple RL seeds per node and reporting confidence intervals would strengthen the statistical validity of the reported optima.
    \item \textbf{Online surrogate refinement:} Jointly training the surrogate model during RL exploration, rather than using a fixed approximation, may improve sample efficiency as the policy converges.
\end{itemize}

\section{Conclusion}

We presented an RL-driven approach to model-specific ASIC design that jointly optimizes compute architecture, memory hierarchy, and workload partitioning for AI inference silicon across \ProcessNodeRange{}. The core contribution is a single MDP formulation with mixed discrete-continuous actions that replaces multi-stage manual RTL iteration with an automated architecture search, producing tape-out-ready configurations directly from the target neural network.

We validate on two workloads spanning distinct optimization regimes. On Llama 3.1 8B FP16 in high-performance mode, the best configuration is \BestNode{} with mesh \BestMesh{} (\BestCores{} active TCCs), achieving \BestTokS{} tokens/s at \BestPowerW{}\,W within \BestAreaMMtwo{}\,mm$^2$. Compared to the worst node (\WorstNode{}), the best delivers \PerfRatio{}$\times$ higher throughput, \AreaReductionFactor{}$\times$ smaller area, and \PPAImprovementFactor{}$\times$ better PPA. On SmolVLM in low-power mode, all 7 nodes achieve $<$13\,mW at 10\,MHz with compact meshes (2$\times$4 at 3nm to 3$\times$4 at 28nm), demonstrating the same RL formulation adapts to power-constrained vision-language workloads. Across \ProcessNodeRange{}, the optimization interface remains stable without node-specific manual retuning.

The formal algorithm (Algorithm~\ref{alg:rl_loop}), quantitative scaling-law fits (Table~\ref{tab:statistical_analysis}), and the end-to-end pipeline (Figure~\ref{fig:pipeline}) provide a reproducible foundation for extending this approach to non-mesh topologies, broader model families, and multi-seed evaluation protocols.

\FloatBarrier

\end{document}